\documentclass[useAMS,usenatbib,usegraphicx]{mn2e}

\title{What planetary nebulae tell us about helium and the CNO
  elements in Galactic bulge stars}
\author[J. F. Buell]{J. F. Buell$^{1}$\thanks{E-mail:
    buelljf@alfredstate.edu}\\ $^{1}$100 College Drive, SUNY College
  of Technology at Alfred, Alfred, NY 14843, USA\\ }
\begin{document}

\maketitle

\date{Accepted 2012 October 15. Received 2012 October 12; in
  original form 2012 June 18}

\pagerange{\pageref{firstpage}--\pageref{lastpage}} \pubyear{2012}

\label{firstpage}

\begin{abstract}
Thermally pulsing asymptotic giant branch (TP-AGB) models of bulge
stars are calculated using a synthetic model. The goal is to infer
typical progenitor masses and compositions by reproducing the typical
chemical composition and central star masses of planetary nebulae
(PNe) in the Galactic bulge. The AGB tip luminosity and the
observation that the observed lack of bright carbon stars in the bulge
are matched by the models.

Five sets of galactic bulge PNe were analyzed to find typical
abundances and central star of planetary nebulae (CSPN) masses. These
global parameters were matched by the AGB models. These sets are shown
to be consistent with the most massive CSPN having the largest
abundances of helium and heavy elements. The CSPN masses of the most
helium rich (He/H$\ga$0.130 or $Y\ga0.34$) PNe are estimated to be
between 0.58 and 0.62$\,{\rm M}_{\sun}$. The oxygen abundance in form
$\log{\rm (O/H)}+12$ of these highest mass CSPN is estimated to be
$\approx$8.85.

TP-AGB models with ZAMS masses between 1.2 and 1.8$\,{\rm M}_{\sun}$
with $Y_{\rm ZAMS}\approx0.31-0.33$ and $Z_{\rm ZAMS}\approx0.19-0.22$
fit the typical global parameters, mass, and abundances of the highest
mass CSPN. The inferred ZAMS helium abundance of the most metal
enriched stars implies $dY/dZ\sim4$ for the Galactic bulge. These
models produce no bright carbon stars in agreement with observations
of the bulge. These models produce an AGB tip luminosity for the bulge
in agreement with the observations. These models suggest the youngest
main sequence stars in the Galactic bulge have enhanced helium
abundance ($Y\approx0.32$) on the main sequence and their ages are
between 2 and 4 Gyrs.

The chemical evolution of nitrogen in the Galactic bulge inferred from
the models is consistent with the cosmic evolution inferred from HII
regions and unevolved stars. The inferred ZAMS N/O ratio ($\log{\rm
  N/O}\approx-0.35$) of bulge PNe with the largest CSPN masses are
shown to be above the solar ratio. The inferred ZAMS N/O ratios of the
entire range of PNe metallicities is consistent with both primary and
secondary production of nitrogen contributing to the chemical
evolution of nitrogen in the Galactic bulge.

The inferred ZAMS value of C/O is less than 1. This indicates the mass
of the PNe progenitors are low enough (M$\la1.8\,{\rm M}_{\sun}$) to
not produce carbon stars via the third dredge-up.

\end{abstract}


\section{Introduction}
Galactic bulge planetary nebulae (GB-PNe) are an important set of
stars for the study of the stellar and chemical evolution of
stars. These important stars give information about the ages of
different populations of stars. The central star planetary nebulae
(CSPN) mass should be related to the zero-age main sequence (ZAMS)
mass of the progenitors. PNe allow direct measurements of the
abundances of important atoms which are difficult to directly measure
in other stars such as helium, neon and argon.

An important unsettled question about the bulge is whether it contains
both old and young populations of stars or just old populations? There
are two contradictory lines of evidence about this
question. Photometric studies of the luminosity of the galactic bulge
main sequence turn-off (MSTO) indicate it is faint
\citep{kr02,zo03,br10,cl11}. This suggests most star formation in the
bulge ended around $\sim10\,{\rm Gyr}$ ago. This view has been
challenged by the work of \citet{ben10, ben11} using spectroscopy of
stars during microlensing events. They found the position of Galactic
bulge stars on the $\log{g}-\log{T_{eff}}$ plane near the MSTO and on
the subgiant branch. In the same study the abundances of several
elements were measured allowing a determination of chemistry. This is
important since the positions of isochrones vary by [Fe/H] and
[$\alpha$/Fe]. To fit these stars with the correct metallicity
isochrones, much younger isochrones then those used in the photometric
studies were needed for a proper fit. The youngest Galactic bulge
stars in \citet{ben11} have inferred ages of $\sim3.0\,{\rm
  Gyr}$. This suggests, in contradiction of the photometry studies,
there exists a significant population of younger stars in the bulge.

\citet{ng12} suggested a way to reconcile these two disparate
interpretations. They suggested that there is a population which is
older than that suggested by \citet{ben10, ben11} but younger than
10$\,{\rm Gyr}$ suggested by photometric studies. They argue such a
population reconciles these two disparate observations with
essentially two old populations; one with enhanced values of $Y$ and
one with normal values of $Y$. They argue that both the photometric
and spectroscopic studies use of non-helium enhanced isochrones give
the wrong inferred ages and masses of bulge stars. Their argument is
that if helium enhanced isochrones are used for comparison to the MSTO
in the photometric studies, a younger age would be indicated. If
helium enhanced isochrones are used to compare to the results from the
spectroscopic studies of \citep{ben10,ben11} older ages would be
indicated. This would bring these contradictory results at least
nearer to agreement.

\citet{na11} also found indirect evidence that there are stars with
enhanced $Y$ in the bulge. They attributed the anomalous Galactic bulge
red giant branch bump to a population of stars with an enhanced value
of $Y$ ($Y\sim0.35$).

An enhanced value of $Y$ in ZAMS stars is not unprecedented. Globular
clusters in the Milky Way and the Magellanic Clouds show evidence of
the existence of multiple populations with distinct values of $Y$. The
extensive spectroscopic evidence is reviewed in \citet{gratreview} and
the photometric evidence of multiple main-sequences, sub-giant
branches, red giant branches and horizontal branches in the same
globular cluster is reviewed in \citet{sqrev09}. In many (and possibly
all) globular clusters there are at least two chemically distinct
populations. The older population (primary) consists of stars with
scaled-solar abundances and a helium abundance which can be determined
by linearly interpolating in $Z$ from the primordial helium abundace,
$Y_0$, to the solar helium abundance, $Y_{\sun}$. The slightly younger
population (secondary) consists of stars with an enhanced abundance of
helium ($\Delta Y$) which can be very modest ($\approx0.01$) or large
($\approx0.12$). The enhancement of helium is defined as the increase
of the helium fraction over what would be determined from the
equation:
\begin{equation}
Y=Y_0+(dY/dZ)Z
\end{equation}
where $Z$ is the metallicity.

There are several available abundance studies of Galactic bulge PNe
(hereinafter GB-PNe) which measure the abundance of helium
\citep{rat92,rat97,cui00,liu01,esc04,ex04,wl07,chi09}. All of these
studies show some GB-PNe have an elevated He/H (Defined here as
He/H$>$0.120). In each of these studies the highest level of He/H for
each lies between 0.135 and 0.22. For a PN in the disc or other region
where there is active star formation an elevated He/H would be
interpreted as the result of helium enhanced material being mixed up
from the interior to the surface of the star. The amount of this
enhancement is greatest for intermediate-mass stars (defined as
M$\ga3.5\,{\rm M}_{\sun}$). The surface helium abundance of
intermediate-mass stars is increased by the action of the second
dredge-up (SDU), the third dredge-up (TDU) and the action of
hot-bottom burning (HBB). HBB occurs at the base of the convective
envelope of a thermally pulsing asymptotic giant branch (TP-AGB) star
when the temperature at the base becomes high enough for the CN cycle
to operate. HBB is limited to intermediate-mass stars ($\ga3.5\,{\rm
  M}_{\sun}$) and converts C to N while producing some He. The SDU
occurs when the star enters the early-AGB (E-AGB) and material which
experienced complete hydrogen buring is mixed up to the surface. A
result of SDU is to raise the surface abundance of
helium. Theoretically to get a SDU a star needs a ZAMS mass of
$\ga4.0{\rm M}_{\sun}$ \citep{bs99}. The TDU occurs near the end of a
helium shell flash when during the TP-AGB the convective envelope
penetrates into a region where partial helium burning has taken
place. However, to significantly enhance the abundance of helium and
nitrogen requires an intermediate-mass star (M$\ga3.5\,{\rm
  M}_{\sun}$. The lifetime of such a star is short ($\la0.5\,{\rm
  Gyr}$).

If intermediate-mass stars are the progenitors of the high helium
bulge PNe then they would come from a very young population. This
hypothetical very young population would be younger than the results
of \citet{ben10,ben11} would indicate. Since a significant number of
the bulge PNe have high helium abundances generally associated with
intermediate-mass stars, but there are not significant numbers of
intermediate-mass stars in the bulge, there must be an alternative
explanation of the observed high helium abundances. The observed range
of GB-PNe He/H is 0.09-0.20. Assuming the progenitor has near solar
abundances, $Y\approx0.27$ and $Z\approx0.017$ \citep{vsp12}, to
produce a helium abundance this high requires a star of mass of
$\ga4\,{\rm M}_{\sun}$ (\cite{gj93,mar99}) with a lifetime of
$\la0.4\,{\rm Gyr}$. Main sequence stars of this mass are not observed
in the bulge. It seems unlikely there are stars of this mass in the
bulge since it would mean there would be AGB stars brighter than seen
in the bulge. If intermediate-mass stars exist in the bulge, it would
mean main sequence stars with masses between 2 and 4$\,{\rm M}_{\sun}$
would also exist in the bulge. These 2-4$\,{\rm M}_{\sun}$ would
produce bright carbon stars. This contradicts the observation that the
C-stars in the bulge are faint. Faint carbon stars are interpreted as
carbon-enriched low-mass stars resulting from a binary mass-transfer
event.

The alternative is that the progenitors of these PNe are older
low-mass stars which formed with high helium abundance. The PNe
abundances would then reflect the initial abundances with minor
modifications due to the first dredge up (FDU) only. The FDU effects
stars of all mass and occurs when the star enters the red giant branch
(RGB). During the FDU the abundances of He and N are slightly
increased at the expense of a decrease in the C
abundance. \citet{bu12} (hereinafter Paper I) presented a similar
model to explain the two PNe in globular clusters (JaFu 1 and JaFu 2)
both of which have elevated helium abundances but are clearly from an
old, low-mass population. Paper I showed the global parameters of both
the PNe (helium abundance, oxygen abundance, mass of central star) and
the parameters inferred from the host cluster (metallicity and
progenitor mass) can be matched by low-mass, helium enhanced models.

The goal of this paper is to test if the GB-PNe can be explained with
a low-mass model and to use their observed global parameters to infer
the ZAMS mass and the ZAMS $Y$ value. In Section~\ref{sec:mod} the
TP-AGB models are discussed. In Section~\ref{sec:abun} the abundances
and CSPN masses of GB-PNe as well as other relevant observations are
reviewed with the goal of determining appropriate global averages to
compare to the models. In Section~\ref{sec:res} the models are
presented and compared to the observations. In Section~\ref{sec:diss}
the implications are discussed. In Section~\ref{sec:con} conclusions
and suggestions for further work are presented.

\section{Models}
\label{sec:mod}

To model the evolutionary behavior of the stars the model described in
Paper I is used. Briefly, this model calculates the structure of the
envelope during the interpulse period of the TP-AGB by geting the
luminosity from a core-mass luminosity relation. The effects of
pre-TP-AGB evolution are modeled using fits to published models. Most
of the relevant details of this model are explained in \citet{bu97},
\citet{bue97}, and \citet{gbm}. In paper I the most significant update
to the model was an updated rule for treating the mass-loss during the
red giant branch (RGB) and early-AGB (E-AGB) phases. This is a
significant model input, especially for low-mass ZAMS stars which can
lose an appreciable fraction of their mass during the RGB and
E-AGB. In paper I, particular attention was paid to the effect of an
enhanced helium abundances on mass-loss during these stages. The
mass-loss in these stages was found by integrating the mass-loss rate
formula over the Padova tracks. There is an important point to note,
in paper I the calculation of pre-TP-AGB mass-loss model, as noted by
the referee of that paper, the evolution of the star and the mass-loss
rates are not coupled. The amount of mass-loss found with these
equations are quite reasonable but the reader should be aware of
it. For the masses of the stars modeled here (M$=1.1-1.8\,{\rm
  M}_{\sun}$) any errors in the pre-TP-AGB mass-loss will not have a
large effect on the subsequent evolution of the star.

\subsection{Elemental abundances}
\label{sec:ea}

The solar abundance set used in this paper is from \citet{asp05}. This
set was chosen because the oxygen abundance ($\epsilon{\rm (O)}=8.66$)
is similar to that found in Galactic HII regions such as the Orion
Nebula. A typical value of oxygen for the Orion Nebula is slightly
higher, e.g. $\epsilon{\rm (O)}=8.74$ (e.g. \citealt{sd11}).

Abundances for stars are usually published as in terms of $[{\rm
    Fe/H}]$. At sub-solar $[{\rm Fe/H}]$ the $\alpha$ elements are
known to be enhanced. At $[{\rm Fe/H}]\la-1.0$ the $\alpha$ elements
are enhanced relative to iron by a constant factor. The level of the
$\alpha$ plateau is expressed as
\begin{equation}
[\alpha/{\rm Fe}]=k_1
\end{equation}
where $k_1$ is an adjustable parameter. In this model this parameter
which will be set to approximately +0.4. At a value of $[{\rm
    Fe/H}]\approx-1$ the value of $[\alpha/{\rm Fe}]$ begins to
decrease linearly and reaches a value of zero at $[{\rm Fe/H}]=0$. The
position of this knee in the distribution is given by
\begin{equation}
[{\rm Fe/H}]_{\rm knee}=k_2.
\end{equation}

The mass fraction of helium $Y$ is computing an intermediate scaled
solar value, $Y_{ss}$ is computed by
\begin{equation}
Y_{ss}=Y_0+\frac{dY}{dZ}Z
\end{equation}
where $Y_0$ is the primordial helium mass fraction and $\frac{dY}{dZ}$
is the slope of the relationship between $Y$ and $Z$. The mass
fraction of helium is then enhanced by an adjustable factor $\Delta
Y$. The equation is given by
\begin{equation}
Y=Y_{ss}+\Delta Y.
\end{equation}

Since the element nitrogen shows considerable variation the abundance
of nitrogen can be enhanced by a factor $\Delta{\rm N}$. When nitrogen
is enhanced the value of $Z$ is enhanced by the same amount. The mass
fraction of hydrogen is calculated from
\begin{equation}
X=1-Y-Z.
\end{equation}

\subsection{Red giant mass-loss}
\label{sec:rgbml}

The mass-loss which occurs on the red giant branch (RGB) is very
important for low-mass stars, which in some extreme cases may prevent
the star from even reaching the TP-AGB. Most of the pre-TP-AGB
mass-loss occurs during the RGB. The standard method to determine the
amount of mass-loss is to use Reimers' Law \citep{rei} given by
\begin{equation}
\dot{M}=\eta\frac{LR}{M}
\end{equation}
where L, R and M are the stellar luminosity, radius and mass,
respectively in solar units. However, in this paper the pre-TP-AGB
mass-loss was calculated using an updated and modified version of the
Reimers formula of \citep{sch} given by
\begin{equation}
\dot{M}=\eta\frac{LR}{M}(\frac{T_{\rm eff}}{4000{\rm K}})^{3.5}
(1+\frac{g_{\sun}}{4300g_{\star}})
\end{equation}
where $T_{\rm eff}$ is the effective stellar temperature, $g_{\star}$
is the surface gravity of the star in cgs units. Values of 27400${\rm
  cms^{-2}}$ for $g_{\sun}$ and $8.0\times10^{-14}$ for $\eta$ were
adopted, which are the values recommended by \citep{sch}. This new
mass-loss rule appears to give better results for horizontal branch
masses then the Reimer's rate \citep{sch}.

This mass-loss law was applied to the variable $Y$ stellar evolution
tracks from the Padova stellar evolutionary library
(http://pleadi.pd.astro.it) described in detail in \citet{berta} and
\citet{bertb}. To determine the red giant mass-loss the mass-loss rate
was integrated from the beginning of the red giant branch (encoded in
the Padova files as brgbs) up to the tip of the red giant branch
(encoded as trgb) using the trapezoidal rule. The amount of mass-loss
between time steps in the models is given by
\begin{equation}
\Delta M_{i}=\frac{1}{2}(\dot{m}_{i+1}+\dot{m}_i)(t_{i+1}-t_{i})
\end{equation}
where $t_i$ and $t_{i+1}$ are the model times and $\dot{m}_{i+1}$ and
$\dot{m}_i$ are the mass-loss rates at the corresponding times. The
total mass-loss is determined by summing all of the $\Delta M_{i}$s.

Paper I gives the mass-losses for the RGB and E-AGB for the Padova
models with metallicities of solar ($Z=0.017$) and lower. Since bulge
metallicities are higher (and potentially super-solar) the
mass-losses for the higher than solar metallicities are presented in
this paper.

\begin{figure*}
\includegraphics[width=188mm]{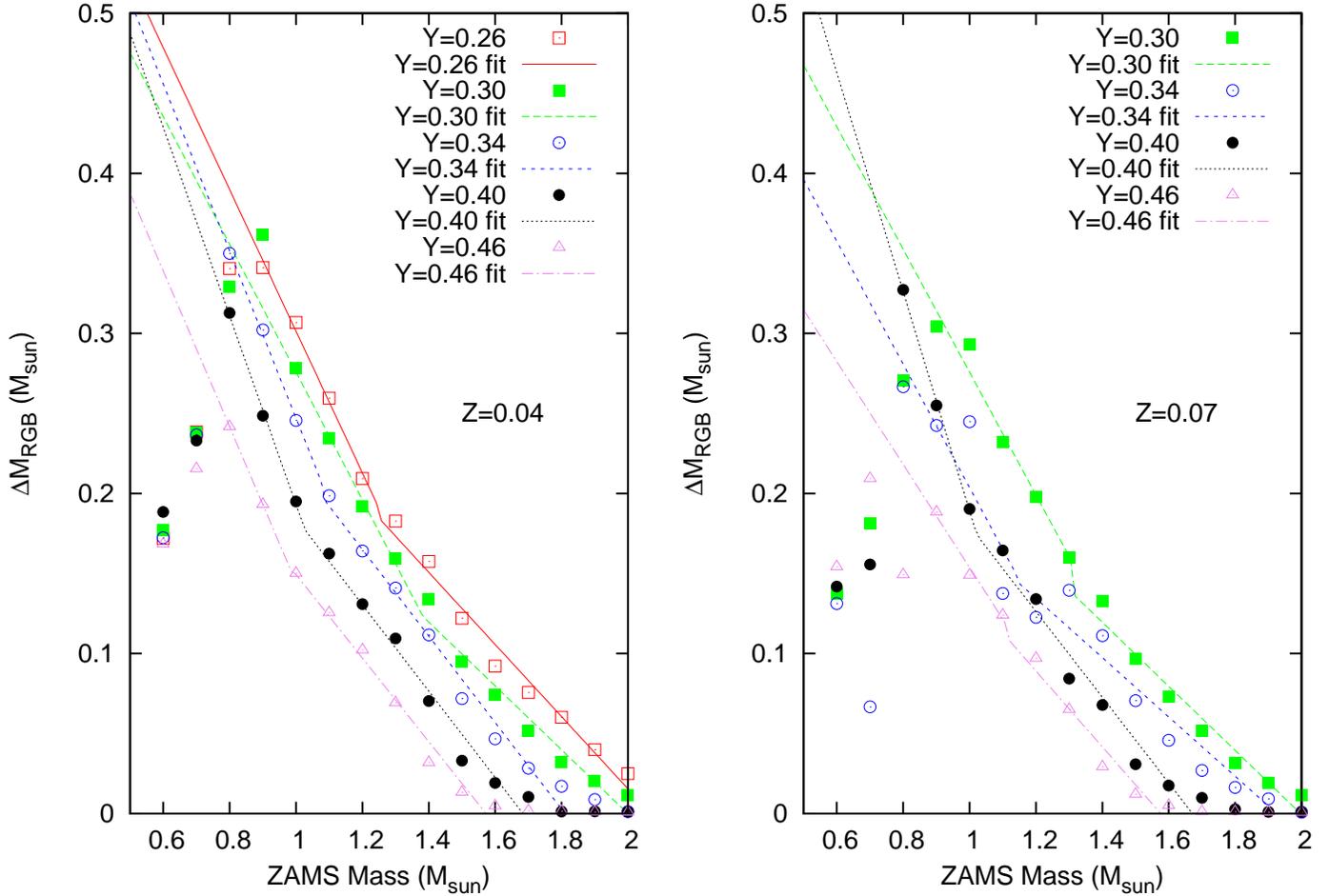}
\caption{Each panel in the figure shows the calculated mass-loss on
  the red giant branch for the $Z$=0.04 and 0.07 for the available
  values of $Y$. The red open squares, green filled squares, blue open
  circles, black closed circles and violet triangles are the
  calculated mass-losses for $Y$=0.26, 0.30, 0.34, 0.40 and 0.46,
  respectively. The red solid, green dashed, blue short dashed, black
  dotted and violet long-dashed dotted lines are the fits for the
  $Y$=0.26, 0.30, 0.34, 0.40, and 0.46 mass-losses, respectively.}
\label{rgbmassloss}
\end{figure*}

The mass-losses on the RGB as a function of the ZAMS mass for all
available values of $Y$ for $Z=0.04$ and $Z=0.07$ are shown in
Figure~\ref{rgbmassloss} by the points. In all panels it is evident
the amount of mass-loss decreases as $Y$ increases. This occurs
because stars with higher values of $Y$ means the RGB star will have
smaller radii and higher surface gravity due to the lower opacity in
the outer layers. These factors lower the mass-loss rates and the
total mass-loss.

The RGB mass-loss was fit using two linear fits for higher and lower
ZAMS masses. The transition point between the fits was determined by
visually estimating the mass where the slope appears to change. This
mass is typically found around a ZAMS mass of 0.8-0.9 M$_{\sun}$. The
higher mass fit was terminated where the high mass line crosses the
horizontal axis. This termination point was estimated visually. For
masses larger than $\approx2.0\,{\rm M}_{\sun}$ the mass-loss is
0. The equations of the linear fits for low and high masses are given
by $\Delta{\rm M}_{\rm RGB,low}$ and $\Delta{\rm M}_{\rm
  RGB,high}$. To make the fits work the fitting was done by excluding
the ${\rm M}\le0.9\,{\rm M}_{\sun}$ models. These can be safely
excluded since the mass-loss for these very low-mass stars eliminates
their envelope before the tip of the RGB is reached and such models
will not be considered in this paper. Visual inspection indicates the
fits are in good agreement to the mass-loss calculations.

The equations for $\Delta{\rm M}_{\rm RGB,low}$ and $\Delta{\rm
  M}_{\rm RGB,high}$ are given by
\begin{equation}
\Delta{\rm M}_{\rm RGB,low}=a_{11}M+a_{10}
\end{equation}
\begin{equation}
\Delta{\rm M}_{\rm RGB,high}=a_{21}M+a_{20}.
\end{equation}
The mass-loss is found by calculating the value of both $\Delta{\rm
  M}$s and finding the maximum value. If the mass-loss is found to be
negative then the value of the mass-loss is set to 0.  The
coefficients of these equations for the different values of $Y$ and
$Z$ are shown in Table~\ref{tab:masslosscoes}. 

\begin{table*}
\caption{Red giant mass-loss coefficients}
\label{tab:masslosscoes}
\begin{tabular}{cccccc}
$Z$&$Y$&$a_{11}$&$a_{10}$&$a_{21}$&$a_{20}$\\\hline
0.040&0.26&-0.442678&0.744023&-0.225394&0.466268\\
0.040&0.30&-0.638065&0.929548&-0.230933&0.455032\\
0.040&0.34&-0.522&0.769042&-0.270372&0.489416\\
0.040&0.40&-0.589165&0.782297&-0.269874&0.454302\\
0.040&0.46&-0.48654&0.630983&-0.26096&0.410264\\
0.070&0.30&-0.3836&0.659316&-0.200032&0.399528\\
0.070&0.34&-0.384918&0.588526&-0.185168&0.356611\\
0.070&0.40&-0.684775&0.873781&-0.267901&0.447723\\
0.070&0.46&-0.3217&0.475414&-0.23702&0.373474\\
\end{tabular}
\end{table*}

A couple of caveats need to be noted. No attempt has been made yet to
calibrate this mass-loss, which will be done in a later
paper. However, the mass-loss values from these equations appear to be
reasonable. For example a $1.0\,{\rm M}_{\sun}$ $Y=0.26$ $Z=0.017$
star would experience 0.28$\,{\rm M}_{\sun}$ of mass-loss on the RGB
which is typical of other models. A typical globular cluster turn-off
mass of $0.80\,{\rm M}_{\sun}$ with $Y=0.245$ and $Z=0.0008$ gives a
RGB mass-loss of 0.22$\,{\rm M}_{\sun}$. This is reasonable since it
gives a zero-age horizontal branch mass of approximately 0.58$\,{\rm
  M}_{\sun}$ which is similar to measured values
(e.g. \citealt{grat10}).

It should be noted, as suggested by the referee of paper I, that the
method used to find the mass-loss is not consistent with the stellar
evolution models. As the star loses mass its surface gravity would
decrease causing the star to expand. For the models used this would
result in a higher mass-loss rate near the tip of the RGB and a
greater amount of mass-loss on the RGB (and the E-AGB) than is
calculated here. However, this effect should be relatively small since
the deviation will only be really significant at the tip of the
RGB. Although the method used here is not strictly consistent the
relative differences in mass-loss due to the effect of the ZAMS helium
abundances and the ZAMS metallicity should be correct.

It was noted by the referee of this paper that \citet{mig12} argued
the RGB mass loss inferred by asteroseismology for the old, metal-rich
($Z\approx0.04$) open cluster NGC6791 is $0.09\pm0.03$.  This is
approximately one half of the red giant mass loss as determined from
the formulas above. The MSTO mass of NGC6791 is 1.23$\,{\rm
  M}_{\sun}$. The amount of RGB mass loss predicted for the
prescription here is $\sim0.17\,{\rm M}_{\sun}$. The relatively small
difference between the mass loss will make little difference to the
subsequent AGB evolution. In the bulge the most massive PN progenitors
will have masses between 1.2 and 1.8$\,{\rm M}_{\sun}$ and only a
relatively small amount of mass ($\la0.1\,{\rm M}_{\sun}$) is lost on
the RGB and E-AGB. Most of the mass loss occurs on the TP-AGB when the
star evolves to the superwind phase. During this phase the mass loss
rates are $10^{-5}$ to $10^{-4}{\rm M}_{\sun}{\rm yr}^{-1}$. In models
when the star reaches the superwind phase with an additional
0.1$\,{\rm M}_{\sun}{\rm yr}^{-1}$ of envelope mass, it produces a
model CSPN with an observationally neglibly different mass. To verify
this some models were run with a pre-AGB mass loss with one half of
the value from the prescription above and the difference in the CSPN
mass is found to be small.

\subsection{Mass-loss on the early-AGB}

The same procedure to determine the mass-loss on the RGB was applied
to the early-AGB (E-AGB) portions of the Padova tracks. An additional
condition of starting the mass-loss when the temperature was below
4500K was assumed since this mass-loss law is applicable only to K and
M stars.

\begin{figure*}
\centering
\includegraphics[width=188mm]{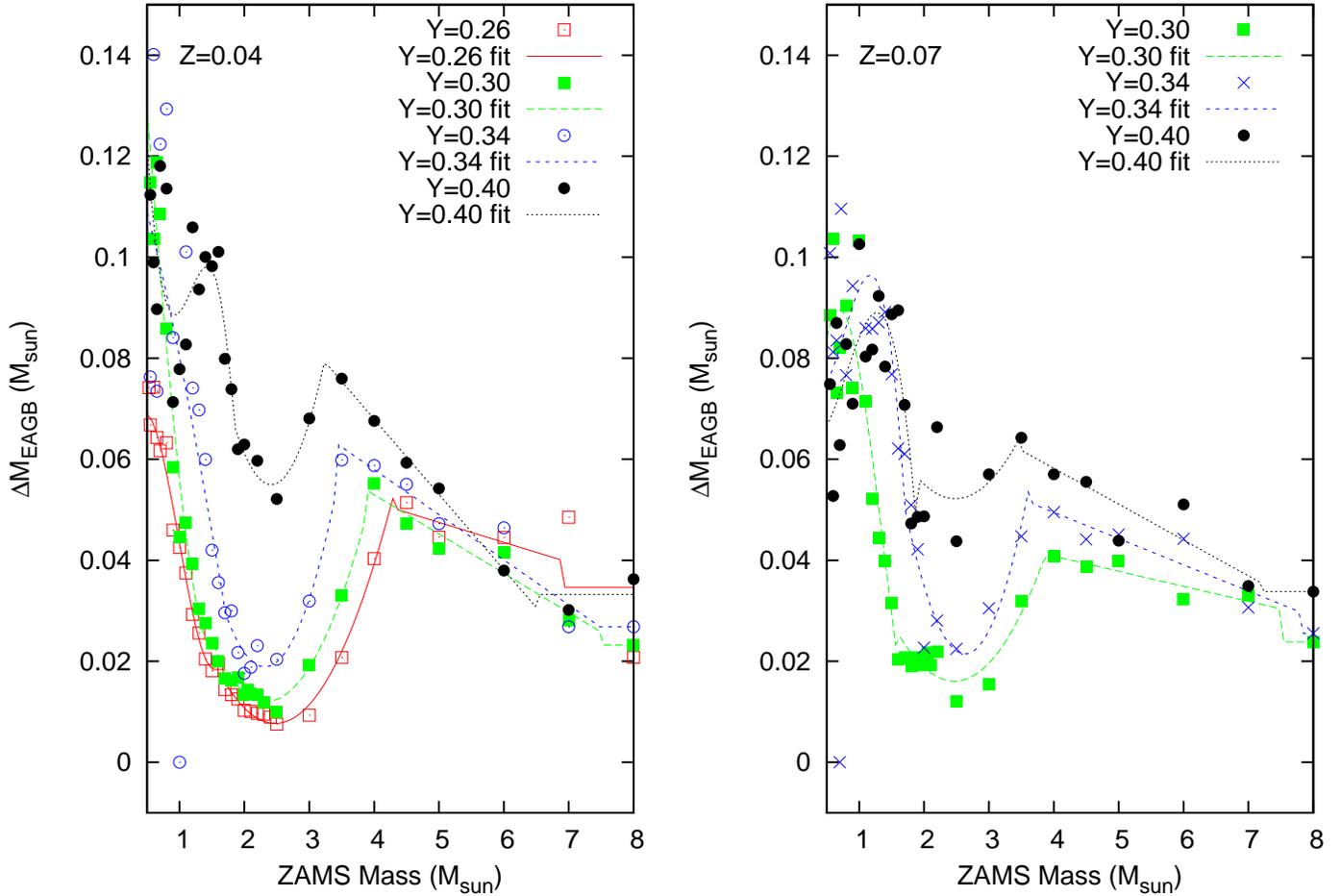}
\caption{Each panel in the figure shows the calculated mass-loss on
  the early asymptotic giant branch as a function of ZAMS mass for the
  $Z=$0.04 and 0.07 models for different
  available values of $Y$. The symbols and lines have the same meaning
  as those in Figure~\ref{rgbmassloss}.}
\label{eagbmassloss}
\end{figure*}

Figure~\ref{eagbmassloss} shows the calculated mass-loss during the
E-AGB and the fits to these mass-losses. The mass-loss on the E-AGB is
fit using 4 fits in different regions of mass. The lowest mass range
(${\rm M}\la1.5\,{\rm M}_{\sun}$) is fit via a cubic, the next mass
range up ($1.5\,{\rm M}_{\sun}\la{\rm M}\la2.0\,{\rm M}_{\sun}$) is
fit using a quadratic fit. The next mass range up ($2.0\,{\rm
  M}_{\sun}\la{\rm M}\la4.5\,{\rm M}_{\sun}$) is fit using a linear
fit. Finally the highest masses are fit using a constant value of
mass-loss. The points of intersection between adjacent fits were
visually estimated. This procedure gives a good fit to the model
mass-losses.

The equations for the E-AGB mass-loss in the first two mass regions
are given by
\begin{equation}
\Delta{\rm M}_{\rm E-AGB}=a_{13}M^3+a_{12}M^2+a_{11}M+a_{10}
\end{equation}
and
\begin{equation}
\Delta{\rm M}_{\rm E-AGB}=a_{22}M^2+a_{21}M+a_{20}
\end{equation}
where $M$ is the mass of the star on the ZAMS. Only the coefficients
of first two regions have been included in Table~\ref{tab:eagbco} to
save space and since no models of sufficient mass which need the fits
for the upper regions are calculated in this paper. The coefficients
for lower metallicities are presented in paper I. The E-AGB mass-loss
is calculated by finding the intersection of the two regions and then
choosing the appropriate region and plugging into the corresponding
equation.

\begin{table*}
\caption{Coefficients for fits to early-AGB mass-loss}
\label{tab:eagbco}
\begin{tabular}{ccccccccc}
$Z$&$Y$&$a_{13}$&$a_{12}$&$a_{11}$&$a_{10}$&$a_{22}$&$a_{21}$&$a_{20}$\\\hline
0.040&0.26&0.0858588&-0.256145&0.182936&0.030799&0.0135521&-0.0670728&0.0906499\\
0.040&0.3&0.0435809&-0.0520068&-0.144506&0.208967&0.0154136&-0.0720782&0.0962957\\
0.040&0.4&-0.156291&0.552081&-0.620921&0.314167&0.0361522&-0.174489&0.265491\\
0.040&0.34&0.00445069&-0.0255256&-0.0266288&0.128011&0.028588&-0.13232&0.171976\\
0.070&0.3&0.0697326&-0.29361&0.306202&-0.00452271&0.0126149&-0.0622221&0.0927601\\
0.070&0.4&-0.0565319&0.135358&-0.0691988&0.0744541&0.0122269&-0.0607111&0.127542\\
0.070&0.34&-0.203047&0.640541&-0.631658&0.273183&0.0337056&-0.177336&0.254678\\
\end{tabular}
\end{table*}

\subsection{Core mass at the first pulse}
\begin{figure*}
\includegraphics[width=168mm]{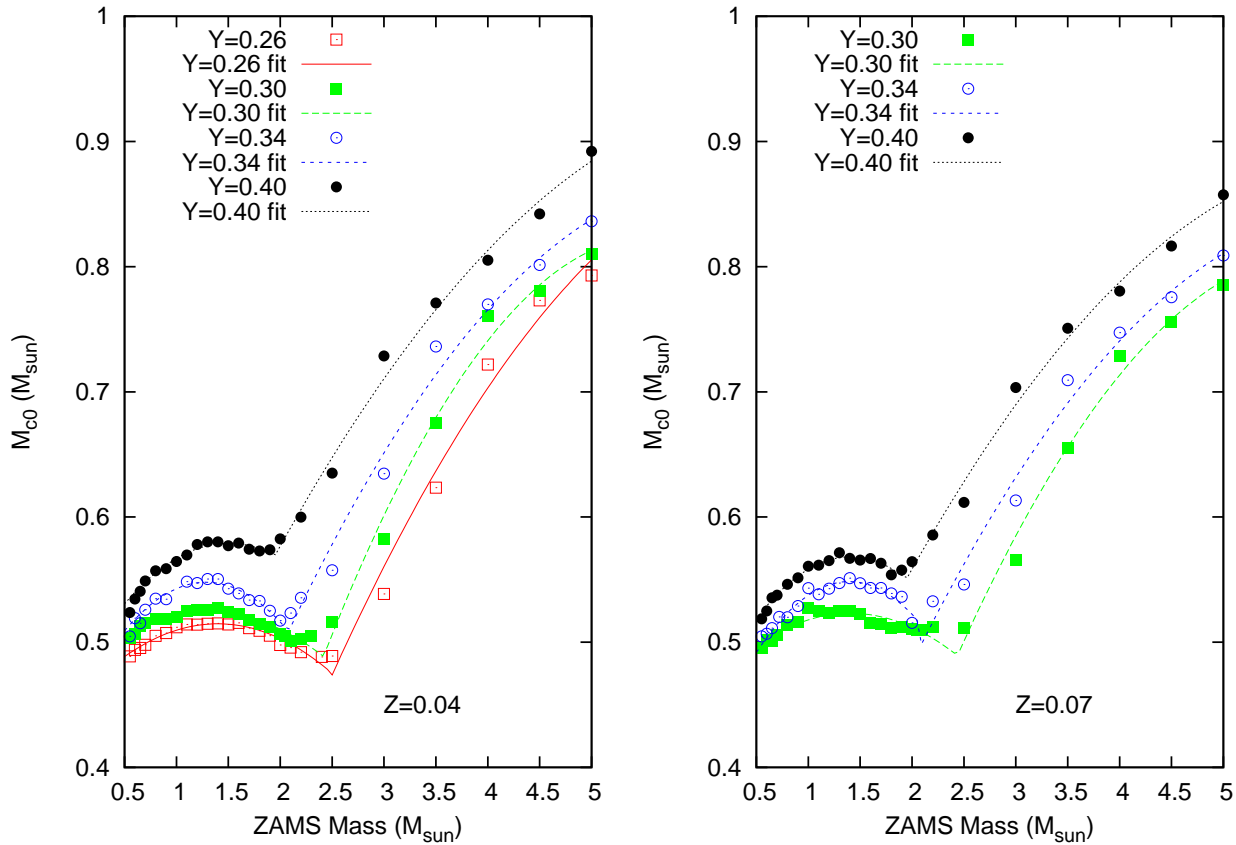}
\caption{The figure shows the core mass at the onset of the first
  pulse for the $Z=$0.04 and 0.07 models and the fits to these
  curves. The symbols have the same meaning as they do in
  Figure~\ref{rgbmassloss}.}
\label{coremass}
\end{figure*}

From the Padova stellar evolution library the core mass at the onset
of the first thermal pulse as a function of the ZAMS mass was
extracted. The tables with this information are found under the
  agb files are found at vizier.u-strasbg.fr in the catalog
  J/A+A/508/355.  Figure~\ref{coremass} shows the mass of the
carbon-oxygen core as a function of mass for several values of $Y$ and
$Z$. An important point to note is that as the initial helium mass
fraction increases so does the mass of the core. The core mass is
important since it is the most important factor controlling the
luminosity of an AGB star. The core mass at the first pulse should
also correlate with the resulting CSPN mass.

Figure~\ref{coremass} show the core mass at the first pulse from the
Padova models for $Z=$0.04 and 0.07 and $Y=$0.26, 0.30, 0.34 and 0.40
as a function of mass. Each set of models with a given $Y$ and $Z$
have been fit by a double quadratic fit, one at lower masses
($\la1.5\,{\rm M}_{\sun}$) and one at higher masses.  The transition
between the two was found between 1.3 and 2.0 ${\rm M}_{\sun}$. The
transition between the lower mass and higher masses was determined by
visual inspection of where the core mass begins to rise steeply. In
all cases the fits to the points and the fit equations are yield core
masses which are typically less than 0.02${\rm M}_{\sun}$ different
from the calculated points.

The equations of the quadratic fits for low and high masses are given
by ${\rm M}_{\rm c0,low}$ and ${\rm M}_{\rm c0,high}$. The equations are
\begin{equation}
{\rm M}_{\rm c0,low}=a_{12}M^2+a_{11}M+a_{10}
\end{equation}
\begin{equation}
{\rm M}_{\rm c0,high}=a_{22}M^2+a_{21}M+a_{20}
\end{equation}
where M is the ZAMS mass. The coefficients for the different values of
$Y$ and $Z$ are shown in Table~\ref{tab:mczero}. The procedure used is to
find the point of intersection between the two fits and then to plug
in the relevant mass.

\begin{table*}
\caption{Coefficients for the fits to the first pulse core masses}
\label{tab:mczero}
\begin{tabular}{cccccccc}
$Z$&$Y$&$a_{12}$&$a_{11}$&$a_{10}$&$a_{22}$&$a_{21}$&$a_{20}$\\\hline
0.040&0.26&-0.0325831&0.0910294&0.451145&-0.0202143&0.284259&-0.1105\\
0.040&0.30&-0.0291444&0.0765474&0.473872&-0.0337214&0.375711&-0.222486\\
0.040&0.34&-0.0625541&0.16101&0.444288&-0.0211523&0.26243&0.0542981\\
0.040&0.40&-0.0483516&0.145099&0.470201&-0.015598&0.211713&0.215733\\
0.070&0.30&-0.0318383&0.0896357&0.45999&-0.0265214&0.314122&-0.118729\\
0.070&0.34&-0.071077&0.198063&0.410435&-0.0196009&0.246253&0.0692979\\
0.070&0.40&-0.0601748&0.169262&0.449156&-0.0167222&0.214947&0.195378\\
\end{tabular}
\end{table*}

The most obvious trend is there is an increase in the core mass as the
value of $Y$ increases. This is important since on the AGB a larger
core mass leads to a higher luminosity. This is also important since
the mass at the first pulse is an important factor in determining the
mass of the CSPN. For a constant ZAMS mass a higher ZAMS $Y$ value
should lead to a higher CSPN mass.

\subsection{Third dredge-up}
In synthetic AGB models the standard method to model the TDU effect is
to use a dredge-up parameter $\lambda$ where
\begin{equation}
\lambda=\frac{\Delta M_{\rm dredge}}{\Delta M_{\rm c}},
\end{equation}
$\Delta M_{\rm dredge}$ is the mass dredged up and $\Delta M_{\rm c}$
is the increase in the core mass during the preceding interpulse
phase. During a thermal pulse the star develops a convective shell in
the intershell region between the base of the hydrogen-rich envelope
and just above the core. This region is helium- and carbon-rich since
it consists of the products of partial helium burning. At the end of
the thermal pulse the convective envelope may penetrate into this
region and mix this carbon- and helium-rich material into the
envelope. The parameter $\lambda$ is a measure of how deeply the
convective envelope penetrates into this intershell region and
determines how much mass is mixed up into the outer layers.

The TDU model used in \citet{gbm} and \citet{bu97} is used. The formula
to calculate $\lambda$ is
\begin{equation}
\lambda=0.90(\log{L_{{\rm He},max}}-\log{L_{{\rm He},min}}).
\end{equation}
The methods to find both $\log{L_{{\rm He},max}}$ and $\log{L_{{\rm
      He},min}}$ can be found in \citet{bu97}. $\log{L_{{\rm
      He},max}}$ is the maximum luminosity of the helium shell during
a shell flash and $\log{L_{{\rm He},min}}$ is the minimum luminosity
required for a dredge-up to occur. $\log{L_{{\rm He},min}}$ is a
function of the mass of the star and its value increases as the mass
gets smaller. The main aspects of this formula are:
\begin{itemize}
\item{Because the maximum luminosity of the helium shell starts below
  the canonical value for a given core mass typically at the first
  pulse this luminosity is too low to produce a dredge-up.}
\item{The value of $\log{L_{{\rm He},min}}$ increases as the mass
    decreases. This prevents TDU in the lowest mass stars.}
\item{For the lower mass models in this paper the TDU makes only a
  minor contribution if it makes any to the abundances of helium and
  carbon. The reason is at most only 1-2 TDU events will occur and the
  values of $\lambda$ will be small ($\sim0.10$).}
\end{itemize}
The abundances of the dredge-up material are determined from the
formulas in \citet{rv81}.

\subsection{Mass-loss on the TP-AGB}

On the TP-AGB, mass-loss is calculated by the pulsation period-mass
loss law of \citet{vw93} without their correction for periods above
500 days. To make the transition from the modified Reimer's rate to
this pulsation mass-loss rule the modified Reimer's rate is used until
the pulsation mass-loss rule becomes larger.

\subsection{TP-AGB models}

The TP-AGB is followed using a synthetic AGB code which is a
descendent of the \citet{rv81} code. The code begins with a guess at
${\rm T}_{\rm eff}$ to calculate the surface boundary conditions. The
equations of stellar structure are then integrated to the base of the
convective envelope. The value of the effective temperature is then
modified until the base of the convective envelope is at the same
position as the core mass. The opacities used for high temperatures
are described in \citet{opal}. For low temperatures the opacities of
\citet{af94} are used. The luminosity during the interpulse phase of
the TP-AGB star is calculated using the expressions in \citet{wg98}. A
mixing length parameter, $\alpha=l/{\rm H}_p$, of 1.70 is used. This
value is chosen since it is close to typical values of values of
$\alpha$ chosen for solar models.

\section{Parameters of the Galactic bulge planetary nebula}
\label{sec:abun}

\subsection{Metallicity, oxygen and helium abundances}
\label{sec:mohe}

\begin{figure*}
\vspace*{74pt}
\includegraphics{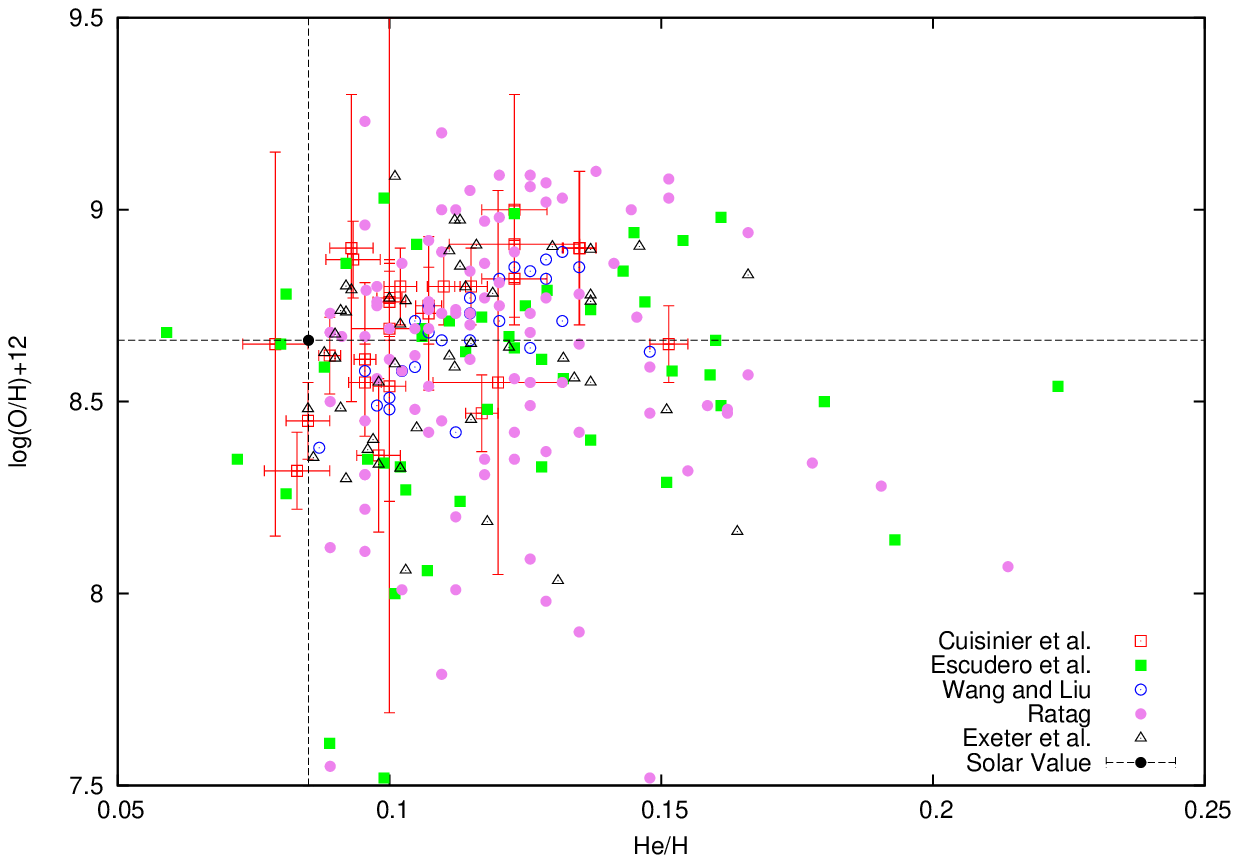}
\caption{The figure shows abundances of log(O/H)+12 as a function of
  He/H. The red open squares, green filled squares, blue open circles,
  violet filled circles and black open triangles are from the GB-PNe
  data from \citep{cui00} (Cuisinier set), \citet{esc04} (Escudero
  set), \citet{liu01} and \citet{wl07} (Wang \& Liu set),
  \citet{rat92,rat97} (Ratag set), and \citet{ex04} (Exeter set),
  respectively. For reference the solar values of \citet{asp05} are
  indicated by the dashed lines.}
\label{fig:bulgeheo}
\end{figure*}

Before trying to match stellar models to the global parameters and
abundances of GB-PNe, it is necessary to decide what values of these
parameters should be matched. Figure~\ref{fig:bulgeheo} shows
log(O/H)+12 as a function of He/H for five different GB-PNe abundance
data sets: the set of \citet{cui00} (Cuisinier set), the set of
\citet{esc04} (Escodero set), the set of \citet{ex04} (Exeter set),
the set in \citet{liu01} and \citet{wl07} (Wang \& Liu set), and the
set in \citet{rat92} and \citet{rat97} (Ratag set). For all sets the
abundances of all elements except He/H are the values reported by the
authors of their analysis of collisionally excited lines (CELs). The
Wang \& Liu papers also report heavy element abundances from optical
recombination lines (ORLs).  These are not used since the abundances
from ORLs are different from the abundances from CELs and are
generally considered to be less reliable (A review of the reliability
and the origin of ORLs can be found in \citep{liu06}.). For all of the
sets the helium abundances are the values reported by the authors from
ORLs.

Overall each of the data sets seem to agree with each other, however,
there are some important differences. Most of the points from the
different sets seem to fall in the same area on the graph. For He/H
values between $\sim0.08$ and $\sim0.14$ the values of O/H for all of
the sets overlap although some of the sets have a larger scatter. The
Escudero and Ratag sets have the largest scatter in O/H and the
Cuisinier and Wang \& Liu sets have the smallest. The biggest
difference is both the Escudero and Ratag sets have a tail of GB-PNe
with He/H greater than 0.16, whereas, the other sets do not.

What is the origin of the high He/H tail in Figure~\ref{fig:bulgeheo}?
In this figure for He/H$\ga0.150$ the value of O/H decreases from a
maximum and appears to level out into a tail. This tail is composed of
PNe from only two of the sets, the Escudero and Ratag sets. Since this
tail does not show up in the other sets, this suggests the most likely
explanation is due to either differences in how PNe were selected or
how the abundances are computed. The abundances for all five sets were
computed using similar sets of ionization correction factors and they
produce very similar results in the He/H range of 0.08 to 0.14. This
suggests the differences arising from the methods of calculating the
abundances are small. This does not rule out errors in computing
  He/H as an explanation. A possible origin of this tail is in the
GB-PNe selection criteria. The different sets use different methods to
determine what is a likely GB-PNe. For example for the Cuisinier set
the criteria for a PN to be in the bulge are (i) its diameter must be
smaller than 10 arcsec, (ii) it must be within 5 degrees of the
galactic centre, and (iii) the 6cm radio flux must be less than 100
mJy. The Escudero set selected for GB-PNe as PNe within 5 kpc of the
centre. This is a less discriminating criteria and as a result a
larger fraction of the Escudero set being disc PNe. \citet{esc04}
estimates $\sim20$ percent of their sample is disc PNe. A possible
explanation of the tail is it consists of inner disc PNe from
intermediate-mass progenitors. For purposes of modeling it will be
assumed this tail is not real since it should show up in all five
sets.

There are a couple of possibilities if this tail is real bulge
population. One possibility is it is produced by low-mass progenitors
with $Y\sim0.4$. This is not unprecedented since some globular
clusters may contain a third, even more helium rich (extreme)
population. Another alternative is these PNe would be the progeny of
galactic bulge blue straggler stars \citep{cl11}. The blue stragglers
in the bulge with an assumed MSTO at 1.4$\,{\rm M}_{\sun}$ could be as
massive as 2.8$\,{\rm M}_{\sun}$. Stars of that mass would experience
many TDU events which would enhance the abundances of helium and
carbon; this could explain the high He abundances. Blue straggler
models for the tail would probably also require enhanced helium
abundances since material from the interior of the star being consumed
would end up near the surface.

\begin{figure*}
\vspace*{74pt}
\includegraphics{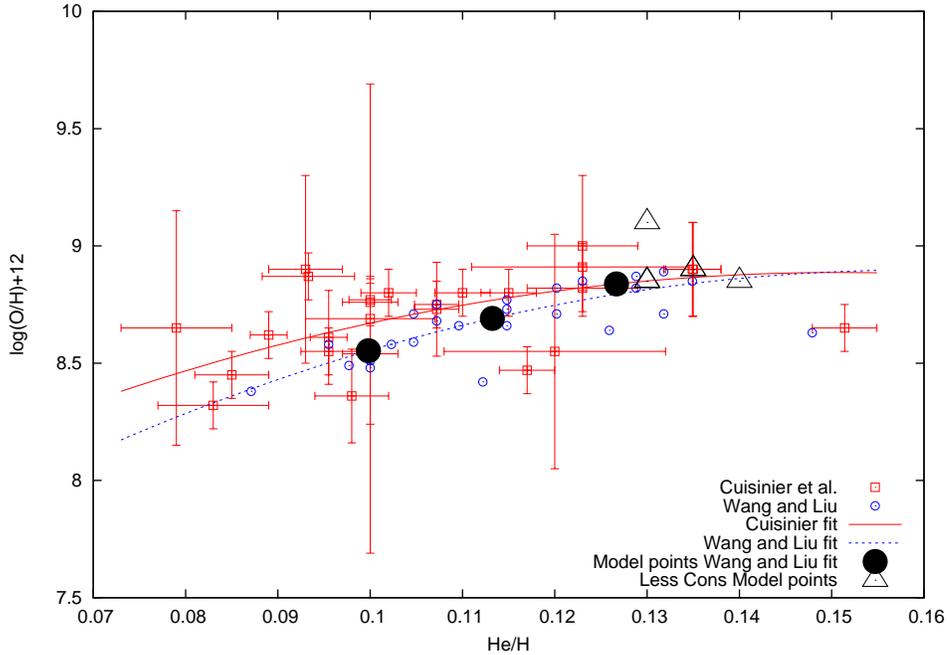}
\caption{This figure is the same as Figure~\ref{fig:bulgeheo} but with
  only the Cuisinier and Wang \& Liu sets only. The solid and dotted
  lines are respectively the fits to the Cuisinier and Wang \& Liu
  set. The large filled circles are the parameters fit by TP-AGB stars
  modeled to the Wang \& Liu fits in~\ref{sec:wl}. The large open
  triangles are the positions of fits of additional models
  in~\ref{sec:lc},~\ref{sec:ml}, and~\ref{sec:ho}. See the text in
  Section~\ref{sec:res} for an explanation.}
\label{fig:bulgeheolimited}
\end{figure*}

Is there a relationship between O/H, which stands in for metallicity,
$Z$, and He/H which stands in for the helium mass fraction, $Y$? In
Figure~\ref{fig:bulgeheo}, by visual inspection, it appears that as
O/H increases so does He/H in the range of He/H of 0.08-0.14. This
trend is much clearer in Figure~\ref{fig:bulgeheolimited} where only
the sets with the smallest scatter, the Cuisinier set and Wang \& Liu
sets, are plotted. In Figure~\ref{fig:bulgeheolimited} there is a
clear trend that as O/H gets larger so does He/H in the He/H range
from 0.08 to 0.135. The Escudero and Exeter sets do not show as a
clear relationship between O/H and He/H as the Cuisinier and Wang and
Liu sets. However, both sets are consistent with there being a
relationship between O/H and He/H, as there appears to be an upward
trend in both. The Ratag sample is noisier then the other samples,
but it is consistent with there being a relationship between O/H and
He/H.

An important point to note is there may be a systematic difference
between the abundances of oxygen as determined from unevolved stars
and PNe. \citet{chi09} compared the oxygen abundances of GB-PNe
determined from CELs to the oxygen abundances determined from giants
and found the PNe abundances are systematically lower by 0.3dex. The
origin of this shift is not clear since dust should, at most, decrease
the abundance by $\sim0.1$dex. It is not clear if this shift is real
or not so the models in~\ref{sec:ho} will look at both possibilities.

Does O/H actually trace O/H of the progenitor star? The abundance of
oxygen in a PN should be close to but not identical to the progenitor
ZAMS abundance, since oxygen should experience a slight decrease
(typically a few percent) in abundance during the FDU at low
masses. At higher masses, which we will not look at in this paper,
oxygen is decreased by the SDU and HBB. Oxygen can be increased in TDU
events. However, all of these changes cause only minor changes in the
abundance of oxygen, so the oxygen abundance in PNe should be a good
indicator of the ZAMS oxygen abundance even with high mass
progenitors.

\begin{figure*}
\vspace*{74pt}
\includegraphics{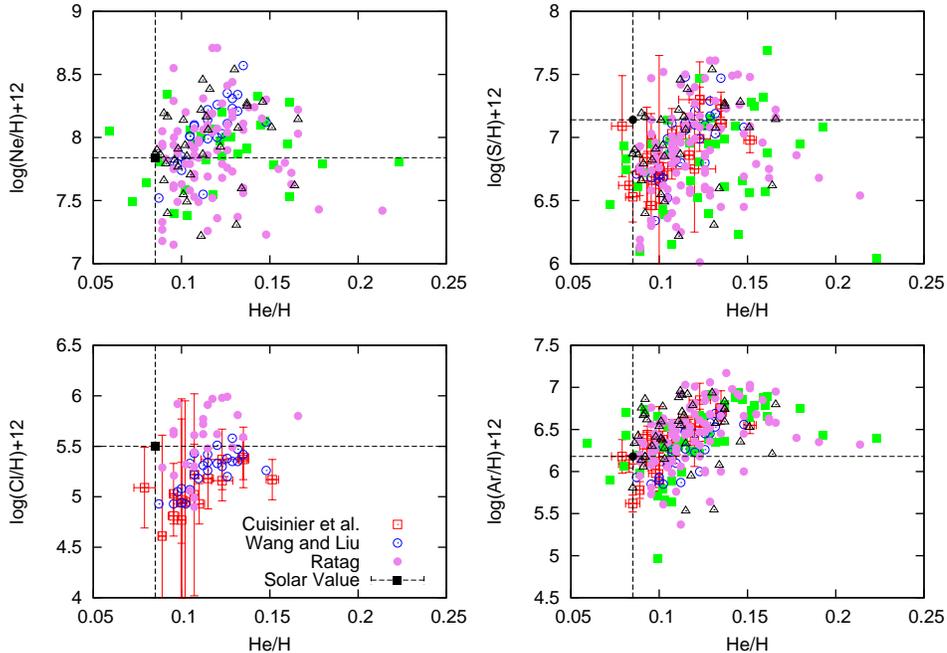}
\caption{The figure shows abundances of log(Ne/H)+12, log(S/H)+12,
  log(Cl/H)+12, and log(Ar/H)+12 as a function of He/H. The symbols
  have the same meaning as they do in Figure~\ref{fig:bulgeheo}.}
\label{fig:bulgehear}
\end{figure*}

Additional support for the proposition He/H increases with metallicity
comes from plotting He/H against heavy elements other than
oxygen. Figure~\ref{fig:bulgehear} shows Ne/H, S/H, Cl/H and Ar/H as a
function of He/H. This figure shows clearly for all the sets
abundances of Ne, S, Cl and Ar increase with He/H up to
He/H$\sim0.14$. None of these elements is thought to be significantly
effected by the stellar evolution between ZAMS and the PN stages.
Neon and argon have an additional advantage, because they are noble
gases, thier abundances will not be affected by being drained onto
dust grains, although for the other elements this effect is probably
small. This figure suggests as the metallicity of the star increases
so does the value of He/H. For all four elements the abundances
increases with He/H up to He/H$\sim0.14$. There is a tail at high He/H
where these abundances drop off. This is the same pattern as seen for
the relationship between the abundance of oxygen and He/H. The high
He/H tail show up in each graph where the Escudero and Ratag sets were
plotted. As noted above its existence is questionable.

In this paper it is assumed the most metal rich PNe are due to the
youngest progenitors. These youngest and most massive progenitors seem
to have a PN He/H between 0.130 and 0.140 and the abundance of oxygen
is in the range 8.80-8.90. If we accept the potential systematic
upward shift of the oxygen abundance by 0.3dex then the oxygen
abundance could be approximately 9.10-9.20 for the most metal rich
PNe. In Section~\ref{sec:res} these will be the abundances of the most
massive GB-PNe matched.

Does it matter if there is a injective relationship between
metallicity and He/H or if there is a spread of metallicities for the
same $Y$ value? The answer is no since there are a significant number
of GB-PNe with He/H$\approx0.14$ in all five sets which require some
explanation. The $Y$ values of PNe can be estimated from the following
formulas
\begin{equation}
X+Y+Z=1
\end{equation}
and
\begin{equation}
\frac{Y}{X}=4{\rm He/H}.
\end{equation}
A maximum $Y$ values can be estimated by assuming the GB-PNe with the
highest He/H have $Z=0.02$. In this case for He/H=0.140 the formulas
above give $Y\approx0.35$. Different, but reasonable assumptions about
$Z$ would not appreciably change $Y$. Clearly the GB-PNe
span a significant range of possible $Y$ and $Z$ values.

In this paper it is assumed that each value of the progenitor $Z$
gives a unique value of $Y$. To get a relationship between O/H and
He/H quadratic fits have been made in Figure~\ref{fig:bulgeheolimited}
to the Cuisinier set and the Wang \& Liu set. The fit to the
Cuisinier set is given by
\begin{equation}
log(O/H)+12=-103.981({\rm He/H})^2+32.4351({\rm He/H})+6.35192.
\end{equation}
The fit to the Wang \& Liu set is given by
\begin{equation}
log(O/H)+12=-97.2275({\rm He/H})^2+30.9835({\rm He/H})+6.42915.
\end{equation}
If He/H=0.130 the value of O/H is respectively 8.81 and 8.81,
respectively, for the Cuisinier and the Wang \& Liu sets. At
He/H=0.09, O/H is respectively 8.43 and 8.43 for the Cuisinier and the
Wang \& Liu sets. Some of the models of the PNe global parameters
from models will be made along these fits but some fits will be made
away from these curves.

\subsection{Central star masses of GB-PNe}
\label{sec:cspnmass}

The mass of the central star of a PN (CSPN) is critical for
determining the mass of the progenitor. Both observational
\citep{wei77,wei00,kal05,cat08} and theoretical studies
\citep{dom99,mar01,mch08} indicate there is a link between the mass of
the progenitor and the mass of the resulting white dwarf. The expected
theoretical relationship is qualitatively confirmed by the
observational studies which show the larger the mass of the progenitor
the larger the mass of the resulting white dwarf. Theoretical studies
indicate the initial-final mass relationship (IFMR) should depend on
metallicity. In particular, the smaller $Z$ is the larger the
resulting white dwarf for any given ZAMS mass. Any potential
metallicity relationship is less clear in the observational studies
\citep{cat08} as there is considerable scatter in the observational
IFMR. Other factors such as stellar rotation may also affect the IMFR
\citep{dom96,cat08}.

In Paper I it was predicted there should be a difference in the IFMR
for different ZAMS value of $Y$. Given the same mass and metallicity a
higher value of $Y$ gives a higher core mass at the onset of the first
thermal pulse. Since the core mass at the first pulse is correlated to
the CSPN mass, a higher $Y$ on the ZAMS should lead to a larger CSPN
mass. 

There do not appear to be a large number of recent studies on the CSPN
masses of GB-PNe. The existing studies are consistent with the
expectation the bulge contains mostly low-mass ZAMS
progenitors. \citet{tyl91} used a variety of methods to get the CSPN
mass; this included plotting the Zanstra luminosities of PNe, L(HI)
and L(HeII) as functions of the respective Zanstra temperatures, T(HI)
and T(HeII). These were compared to the theoretical CSPN tracks of
\cite{sch81,sch83} to get the CSPN masses. \citet{tyl91} also compared
the visual magnitudes of the CSPN to the expansion ages and the
temperatures to their $f$ parameter to get CSPN masses. They got an
average mass of the Galactic bulge CSPN of $0.593\pm0.025\,{\rm
  M}_{\sun}$. Their Figures 11 and 12 present the CSPN mass
distribution of all their masses and all the best determined masses,
respectively. There is a peak in both figures of the number of CSPN
masses at $\sim0.58\,{\rm M}_{\sun}$.  The distributions tails off
quickly to $0.65\,{\rm M}_{\sun}$. About 10 percent of their sample
have masses greater than 0.62$\,{\rm M}_{\sun}$ in their sample. This
is nearly the same as their estimate of the number of disc objects
contaminating the sample \citep{st91}. The high mass tail could be due
to sample contamination by foreground objects or could be due to blue
stragglers in the bulge. This study suggests the highest mass CSPNs
have masses between 0.58 and 0.62$\,{\rm M}_{\sun}$.

\citet{ratagphd} determined the luminosities and temperatures by two
methods. The first is a version of the Zanstra method which adds in
the energy from the far infrared part of the spectrum to get the
stellar temperature and the luminosity.  The second is a method using
an photoionization model where important line ratios were matched to
observations to get the spectral distribution of the stellar continuum
The stars' effective temperature is derived from the continuum. The
luminosity in the second method is determined by adding the energy
from all emission lines, the free-free emission and the infrared
luminosity and the portion of the stellar spectrum emitted with
wavelength less than $91.2\,{\rm nm}$. The agreement between the
luminosities and temperatures determined by various methods is pretty
good (See Figure~2 and Figure~4 in \citealt{ratagphd} Chapter 4 for
these comparisons.).

\begin{figure*}
\vspace{54pt*}
\includegraphics{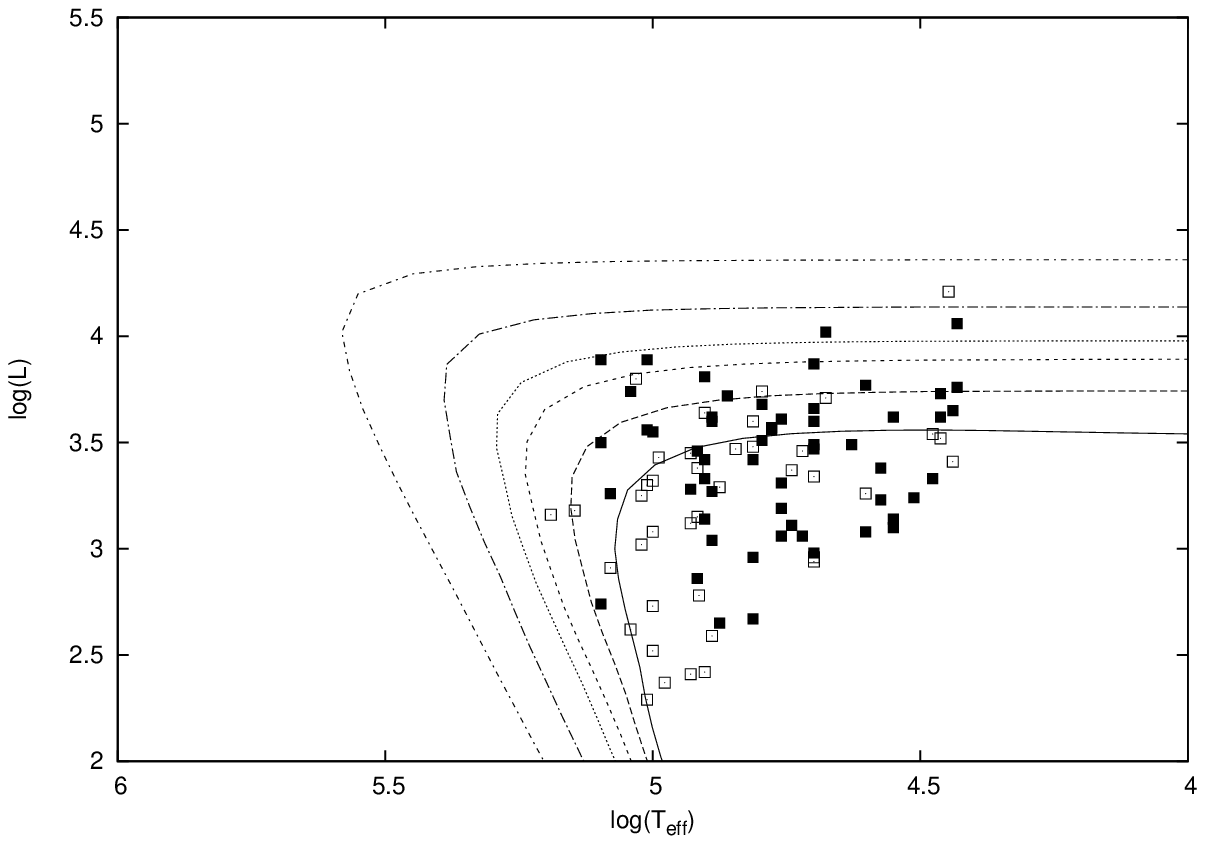}
\caption{This figure shows the positions of the Ratag set on the HR
  diagram with the theoretical \citet{vw94} put in for comparison. The
  filled squares are the PNe with He/H$<0.120$ and the open squares
  are the PNe with He/H$>0.120$. The solid, long dashed, short dashed,
  dotted, long dash-dot, and short dash-dot are the theoretical tracks
  of 0.569, 0.597, 0.633, 0.677, 0.754, and 0.900$\,{\rm
    M}_{\sun}$. All the tracks have ZAMS $Z=0.016$.}
\label{fig:rataghr}
\end{figure*}

\citealt{ratagphd} plotted these luminosities and temperatures on HR
diagrams with theoretical tracks. This figure is partially duplicated
in Figure~\ref{fig:rataghr}. Almost all the Ratag PNe fall between the
$0.546\,{\rm M}_{\sun}$ and the $0.644\,{\rm M}_{\sun}$
tracks. However, the majority falls between $0.546\,{\rm M}_{\sun}$
and $0.600\,{\rm M}_{\sun}$. This suggests the most massive
  GB-PNe CSPN are between 0.58 and 0.60$\,{\rm M}_{\sun}$. This is
  consistent with the \citet{tyl91} results.

Both the \citet{tyl91,ratagphd} results suggest that the progenitors
of the GB-PNe are relatively low-mass since the observed CSPN masses
are low. Both of these studies are quite old and neither has been
repeated recently so some skepticism is justified. However, the
distance to the bulge is relatively well known and the intensity of
the radiation from the GB-PNe can be measured with some
confidence. Many of these PNe appear to fall on the horizontal part of
the CSPN tracks. Since the luminosities, which are nearly constant in
this region, can be fairly well constrained, the CSPN masses should be
reasonably accurate ($\Delta{\rm M}_{\rm CSPN}\sim0.03\,{\rm
  M}_{\sun}$).

Some additional support for a low maximum bulge CSPN masses comes from
the luminosity of the tip of the AGB branch. \citet{zo03} found the
maximum bolometric magnitude of the AGB tip is -5.0 which corresponds
to a luminosity of about $8000\,{\rm L}_{\sun}$
($\log{L}\approx3.9$). In Figure~28 of \citealt{zo03} most of the
stars at the tip of the AGB are closer a bolometric magnitude of -4.5
which corresponds to a luminosity of about $5000\,{\rm L}_{\sun}$
($\log{L}\approx3.7$). If you compare these luminosities to the
horizontal parts of the CSPN tracks of \citet{bl95} this implies a
CSPN mass $\sim0.605\,{\rm M}_{\sun}$ track. Therefore, the AGB tip
luminosity suggests the majority of the CSPNs have masses less than
$0.61\,{\rm M}_{\sun}$.

Putting all of the evidence together suggests the most massive CSPN
PNe in the bulge are between 0.58 and 0.62$\,{\rm
  M}_{\sun}$. Therefore, for the purpose of model fitting in this
paper, it is assumed all the masses of the CSPN of GB-PNe are less
than or equal to $0.620\,{\rm M}_{\sun}$. This maximum mass may be
less than $0.600\,{\rm M}_{\sun}$.

Another study which looked more recently at the question of bulge CSPN
masses is \citet{hul08}. In this study the spectra of the central
stars of a small sample of GB-PNe was obtained. From these the
abundances, the effective temperatures, and the surface gravities were
obtained. The effective temperature and surface gravity were compared
to theoretical tracks to get the CSPN mass. Using this method a higher
average mass is obtained for their small sample (5 objects) of PNe of
$0.696\,{\rm M}_{\sun}$, which is higher than studies which measure
the effective temperature and luminosity using the nebula. Two of the
GB-PNe have a mass of nearly $0.800\,{\rm M}_{\sun}$ which implies
intermediate-mass progenitors (M$\ga3.5\,{\rm M}_{\sun}$). There is
reason to be skeptical of this though. As the authors note, if their
inferred luminosities are used to find the distance to the Galactic
bulge is $10.7\pm1.2\,{\rm kpc}$. This distance which is 25 percent
larger than the typical determination, e.g. $8.4\pm0.4\,{\rm kpc}$
\citep{gh08}.

\begin{figure*}
\vspace{74pt*}
\includegraphics{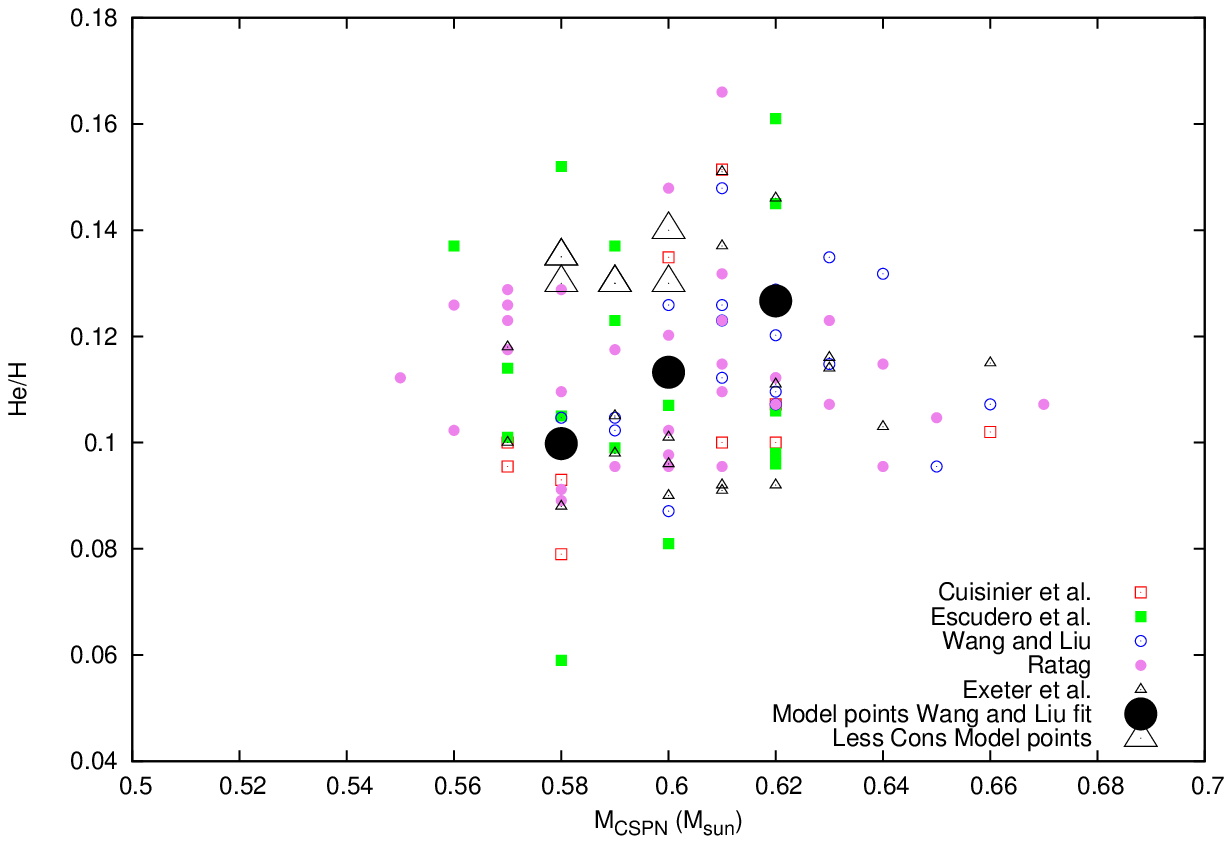}
\caption{The figure shows the value of He/H for GB-PNe versus the
  central star mass from \citet{tyl91}. The symbols have the same
  meaning as in Figure~\ref{fig:bulgeheo}. The large filled circles
  and large open triangles have the same meaning as in
  Figure~\ref{fig:bulgeheo}}
\label{fig:bulgehemc}
\end{figure*}

What is the relationship between the helium abundance and the CSPN
mass of the the GB-PNe? In Figure~\ref{fig:bulgehemc} the five
different samples of GB-PNe He/H are plotted as a function of their
CSPN mass from \citet{tyl91}. Visual inspection of this figure
indicates there is a lot of scatter in all sets but the graph is
consistent with an upward trend between $0.55$ and $0.62\,{\rm
  M}_{\sun}$. If there is an upward trend it appears to end between
0.60 to 0.63$\,{\rm M}_{\sun}$. The lack of a significant correlation
could be due to the relatively large errors expected in the individual
CSPN masses. A positive correlation between He/H and CSPN mass would
be expected if as might be expected in any typical chemical evolution
model where the ISM is being simultaneously enhanced in both helium
and metals. Therefore the more massive CSPN should have higher values
of $Y$ and $Z$.

At even higher core-masses it appears the value of He/H drops off into
a tail with a relatively modest He/H ($\sim0.110$). Is the high CSPN
core mass ($\ga0.62\,{\rm M}_{\sun}$) tail real effect in the bulge or
is it due to disc PNe contamination?  Although this paper can not
definitively answer this, the fact most of the sets have similar He/H
in the tail part of the distribution suggests the tail may be
real. What might cause this tail? If it is real then one possibility
is these are foreground objects misidentified as bulge objects. If
they are disc objects they would appear more luminous and thus more
massive. If they are foreground objects then these PNe would have
values of He/H typical of the disc PNe (He/H$\approx0.110$). Another
possibility is these objects are the progeny of blue straggler star in
the bulge. Potential blue stragglers have recently been identified in
the bulge and the progeny of these more massive stars would be more
massive CSPN.

\begin{figure*}
\vspace{54pt*}
\includegraphics{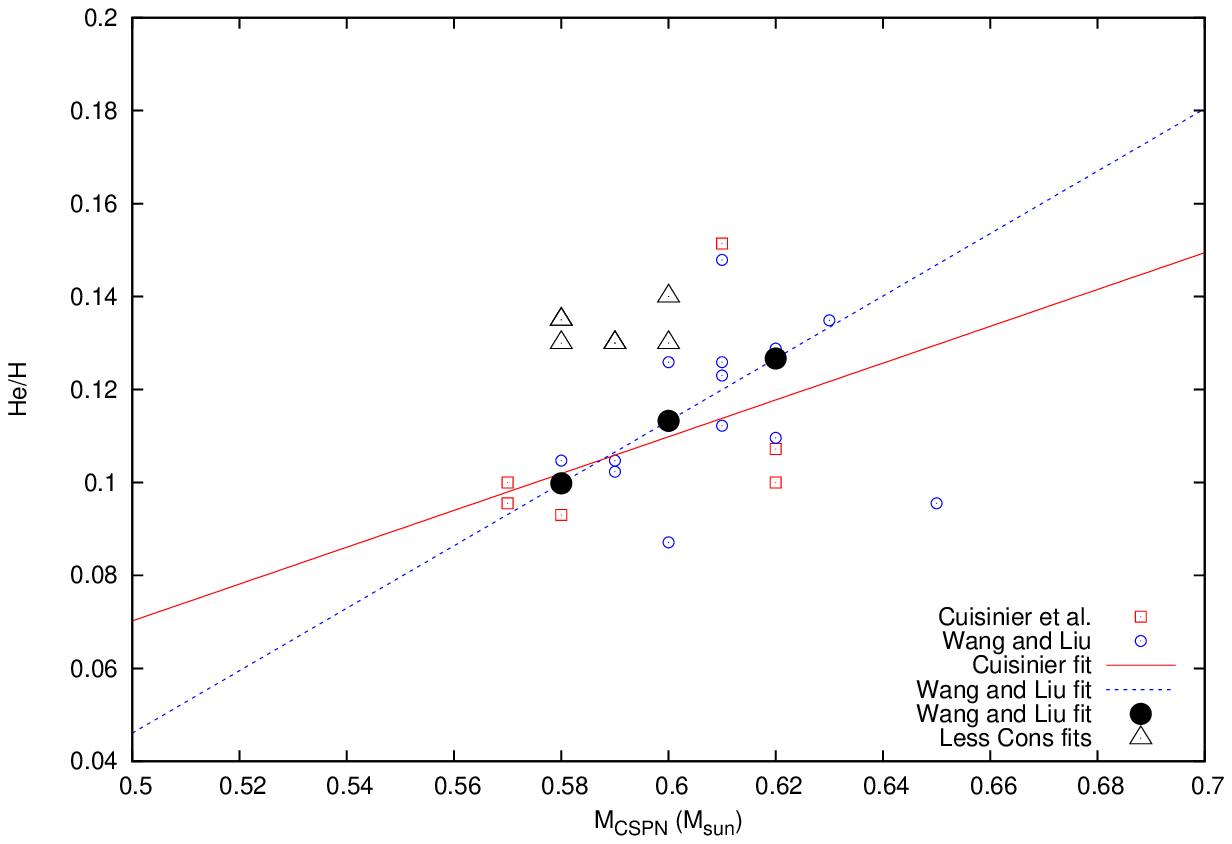}
\caption{The figure is the same as Figure~\ref{fig:bulgehemc} but only
  the Cuisinier and Wang \& Liu sets are plotted against the best CSPN
  masses of \citet{tyl91}. Linear fits to both sets are shown to make
  the trends are clear.}
\label{fig:bulgehemcbest}
\end{figure*}

To get a better look at the relationship between core mass and He/H
from the various GB-PNe of the Escudero set and the Wang \& Liu set
are plotted as a function of CSPN mass, where the CSPN masses are from
the best set from the \citealt{tyl91} paper (as identified in that
paper), in Figure~\ref{fig:bulgehemcbest}. Both of these sets seem to
show as CSPN mass increases so does the value of He/H. This does not
mean there is a unique relationship between CSPN mass and He/H. Only
the fit to the Wang \& Liu set has a truly significant correlation
coefficient. The lack of a definite relationship could be due to the
significant scatter in the CSPN masses.

Although there is possibly no unique relationship between He/H and
CSPN mass this paper will assume there is one. The Wang \& Liu set
seems to have the clearest relationship between He/H and CSPN
mass. The fit to the Wang \& Liu set is given by
\begin{equation}
{\rm He/H}=0.671821({\rm M}_{\rm CSPN}-0.60)+0.113244.
\end{equation} 
This equation will be used to derive the parameters to be fit in
Section~\ref{sec:wl}.

In Figure~\ref{fig:rataghr} the Ratag sample is separated into two
samples with high He/H ($>0.120$) and low He/H ($<0.120$) and plotted
on the HR diagram using the Ratag values of $\log{L}$ and
$\log{T_{eff}}$. This is done since this gives a larger set of CSPN
masses than the \citealt{tyl91} CSPN masses. Visual inspection
  reveals there is at most a minor difference in the distribution on
  the HR diagram of the two subsets. More of the high He/H GB-PNe seem
  to fall on the cooling part of the tracks. This indicates the high
  He/H PNe a slightly higher CSPN mass, since more massive CSPN evolve
  more quickly (e.g. see \citet{vw94}). This makes it more likely more
  massive CSPN will end up on the cooling part of the tracks.

\subsection{Helium versus N/O in GB-PNe}

Nitrogen is a very important element in PNe since it is often used as
an indicator of nuclear processing of material in the interior. This
material is then transported to the surface via convection. In low and
intermediate-mass stars nitrogen is typically enhanced at the FDU and
SDU. Nitrogen can also be enhanced when CNO process occurs at the base
of the convective envelope in AGB stars, known as HBB.

Bulge stars are thought to be fairly low-mass, the work of
\citet{ben10,ben11} suggests the youngest stellar ages are
$\sim3\,{\rm Gyr}$ of age which would suggest a maximum MSTO mass of
$\sim1.5\,{\rm M}_{\sun}$. Stars of this mass should only experience
the FDU which has only a limited effect on N/O, typically doubling
N/O, which is an increase of about 0.3 dex. For a ZAMS star with a
solar ratio of N/O this would mean N/O would be around 1/3 ($\log{\rm
  (N/O)}\approx-0.5$) in a PNe.

\begin{figure*}
\vspace{54pt*}
\includegraphics{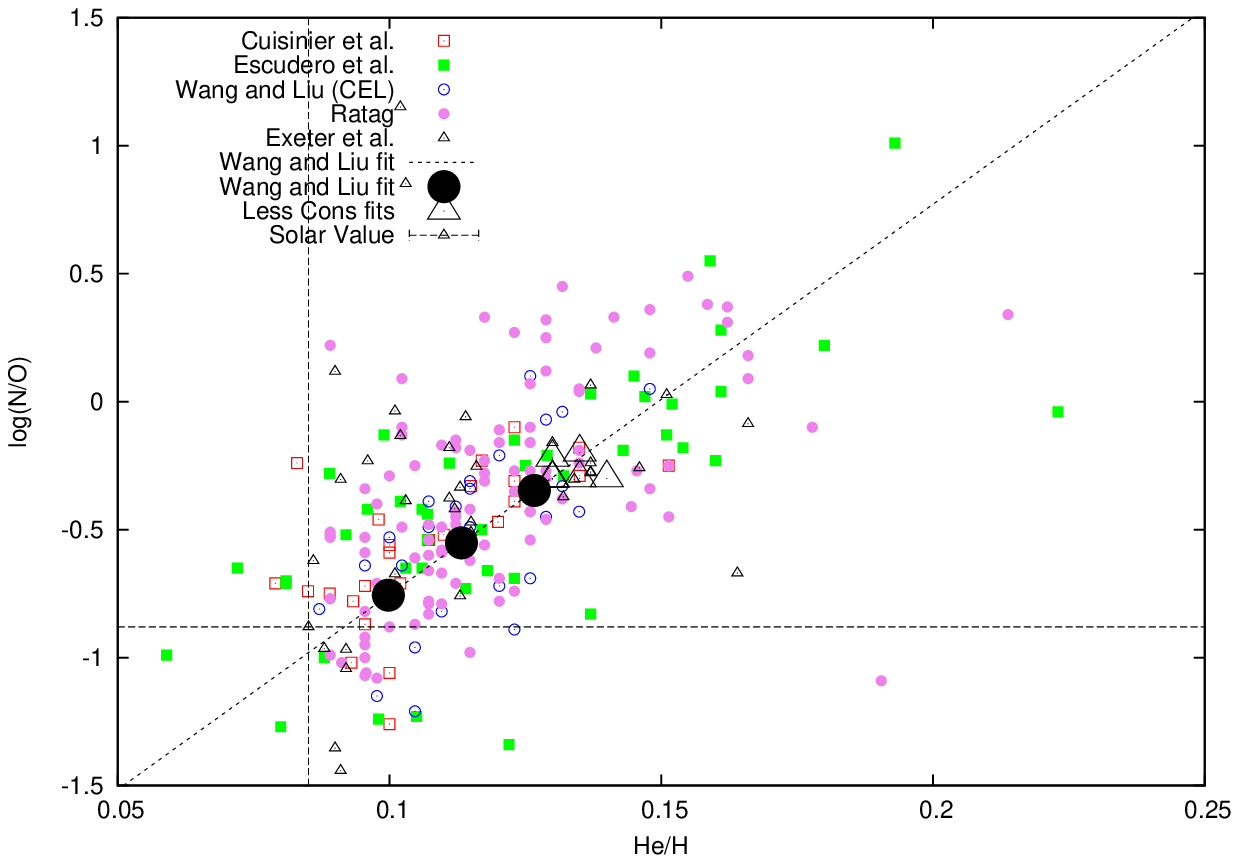}
\caption{This figure shows the positions of the values N/O as a
  function of He/H for the GB-PNe. The symbols have the same meaning
  as in Figures~\ref{fig:bulgeheo} and~\ref{fig:bulgeheolimited}.}
\label{fig:bulgeheno}
\end{figure*}

In Figure~\ref{fig:bulgeheno} the ratio of N/O of GB-PNe is plotted as
a function of the ratio He/H for all five sets. There is clearly a
relationship between N/O and He/H for GB-PNe, with both increasing at
the together in all of the sets. A fit to the Wang \& Liu set, which
is shown on the figure gives:
\begin{equation}
\log{\rm N/O}=15.25({\rm He/H}-0.100)-0.755.
\end{equation}
This line appears to be a good fit to all of the data sets. It will be
adopted to determine the GB-PNe N/O parameter to match in some cases.

Of particular importance is the fact the highest N/O ratios are $\la1$
in almost all of the sets. The Cuisinier and Wang \& Liu sets all of
the GB-PNe have N/O$\la1$. For the Exeter set with the exception of
two clear outliers with N/O$\approx10$, all the GB-PNe N/Os in their
sample are $\la1$. The Escudero set has only a two PNe with N/O
significantly larger than 1. Only the Ratag set has a significant
number of GB-PNe with N/O$\ga1$. For purposes of this paper it will be
assumed for GB-PNe N/O$<1$ since all sets have points in this range
and most of them do not have a significant number of GB-PNe with N/O
significantly larger than 1.

An N/O of $\sim1$ suggests either the progenitors are
intermediate-mass or else are low-mass with a higher than solar N/O
ratio, 0.13. The second is very interesting since it would suggest the
gas from which these stars formed was enhanced in both helium and
nitrogen. This would indicate the bulge ISM was enhanced by the
products of CNO burning.

\subsection{Carbon in GB-PNe}

Another very important element is carbon. Information about this
element for GB-PNe is limited because the bulge is heavily reddened,
and the most reliable lines for determining the carbon abundance in
PNe are in the ultraviolet. The Wang \& Liu set contains a limited
number of measurements of the ratio of C/O from both CELs and
ORLs. Most of their C/O ratios from both methods are below 1 but there
are a small number with ${\rm C/O}>1$ which could indicate TDU
events. This is consistent with the findings of \citet{gu08,pc09,st12}
who found the majority of GB-PNe have oxygen-rich dust spectrum but
there is a minority of stars ($\approx10$ percent) which have
carbon-rich dust spectrum. A large fraction ($\approx40$ percent) of
GB-PNe have a mixed dust spectrum with both carbon and oxygen
features. The mixed features may be associated with a TDU event
occurring near or after the transition from an AGB star to a PN, but
it may also be due to a dense torus around the progenitor where,
during the transition from the AGB to PN phases, chemical reactions
occur in an oxygen rich environment which produces poly-aromatic
hydrocarbons (PAHs). This PAH production shows up as the
carbon-features in a mixed dust spectrum \citep{guz11}. In the second
scenario a mixed feature spectrum does not necessarily indicate a
carbon-rich star in which a TDU occurred.

The conclusion is the majority of progenitors of GB-PNe do not
experience a TDU event. This is supported by the lack of bright carbon
stars in the bulge \citep{alr88}. This indicates the carbon stars in
the bulge which do exist probably formed as the result of a mass
transfer in a binary system and not by the action of the TDU. This
further supports the idea that the stars in the bulge are older,
since only stars with M$\la1.6\,{\rm M}_{\sun}$ do not experience a
TDU.

There may be some indication that a minor amount of TDU
occurs. \citet{utt07} observed Tc in bulge AGB stars which is an
indication of the action of the TDU. In their spectra Tc was only
observed in the brightest and reddest stars with the longest periods
indicating they are probably the most evolved and closest to the
transition to the PN phase. This suggests that the youngest stars in
the bulge may be just massive enough to experience one or two
dredge-ups before entering the PN phase. Depending on the mass of the
residual envelope an AGB star could become a carbon star at the last
moment. This would agree with the small number of GB-PNe with ${\rm
  C/O}>1$. From their luminosities \citealt{utt07} concluded that the
progenitors of these Tc rich stars are $\approx1.5\,{\rm M}_{\sun}$.

All indications are the TDU plays only a minor role in modifying the
abundances in PNe. This is important since the TDU can significantly
modify the abundance of helium but it takes several TDU events to
significantly modify it. Therefore, He/H for these lower mass models
is mostly a function of the initial value of He/H and the result of
the FDU. It will be assumed for fitting purposes, the C/O for GB-PNe
is less than 1.

\section{Results} 
\label{sec:res}

In the previous section it was noted there are a range of possible
GB-PNe parameters to fit. In this section models are calculated which
fit different potential but reasonable parameters of the highest mass
CSPN. The first set of GB-PNe parameters fit in this paper is defined
by the Wang \& Liu fits. This set shows the most well defined trends
and it seems to be a reasonable fit. After that this paper will look
at some less conservative possibilities where the maximum bulge CSPN
mass is less than $0.62\,{\rm M}_{\sun}$.

The basic fitting technique is as follows. A set of typical PNe
parameters to fit is chosen. These parameters are the assumed CSPN
mass and the PNe values of He/H, O/H, and N/O. The ZAMS mass, [Fe/H],
$Y$ and N/O were chosen and a model run. The ZAMS parameters were
modified until a fit to the PNe parameters was acheived.

\subsection{Wang \& Liu set}
\label{sec:wl}

In this subsection models are matched to points on the fits to the
Wang \& Liu set. These fits were chosen since this set shows the most
well defined trends. There is also a defined trend in CSPN mass, which
will be used to explore the chemical evolution of the
bulge. Table~\ref{tab:wangandliu} shows the PNe parameters fit, the
ZAMS parameters of the best fitting models. The positions of the
GB-PNe parameters to be fit are indicated in
Figures~\ref{fig:bulgeheolimited},~\ref{fig:bulgeheno} and
~\ref{fig:bulgehemcbest}. In these fits the most massive CSPN fit have
a mass of $0.62\,{\rm M}_{\sun}$. This probably too high maximum CSPN
mass will produce the largest bulge progenitor mass. The top of the
trend was fixed at a CSPN mass of $0.62\,{\rm M}_{\sun}$ in the fits
to the Wang \& Liu GB-PNe this will produce an He/H $\approx0.130$
which is similar to the maximum He/H in GB-PNe as defined in
Subsection~\ref{sec:mohe}. The lowest mass models have a CSPN mass of
$0.56\,{\rm M}_{\sun}$.



\begin{table*}
\caption{Parameters to fit and parameters of model fit for Wang \& Liu set}
\label{tab:wangandliu}
\begin{tabular}{cccccccccc}\hline
\hline
CSPN mass&He/H&log(O/H)+12&log(N/O)&M&$Y_{ZAMS}$&$Z_{ZAMS}$&[Fe/H]&log(N/O)$_{ZAMS}$\\\hline
0.58&0.09981&8.553&-0.559&1.43&0.257&0.0087&-0.18&-0.800\\
0.60&0.11324&8.691&-0.400&1.63&0.293&0.0122&0.003&-0.630\\
0.62&0.12668&8.794&-0.347&1.80&0.320&0.0175&0.195&-0.573\\\hline
\end{tabular}
\medskip

The CSPN mass, He/H, log(O/H)+12, and log(N/O) are the PN parameters
being fit. M, $Y_{ZAMS}$, $Z_{ZAMS}$,[Fe/H] and N/O$_{ZAMS}$ are the
model ZAMS parameters of the best fitting model. In all of these
  models the parameter $k_1$=[$\alpha$/Fe] is set as +0.35. The end of
  the alpha plateau, $k_2$=[Fe/H], is set to -1.
\end{table*}

To produce the highest mass CSPN a model with $M\approx1.8\,{\rm
  M}_{\sun}$, $Y=0.32$ and $Z=0.0175$ is required. The approximate age
of a 1.8 solar mass $Y=0.32$ and $Z=0.0175$ star is about 2 Gyr. This
is a higher mass and younger age than indicated by the work of
\citet{ben10,ben11}, where their largest mass is $1.25\,{\rm
  M}_{\sun}$ with an inferred age of $2.9\,{\rm Gyr}$. This should be
considered an upper limit on the ZAMS mass.

However, an initial mass of 1.8$\,{\rm M}_{\sun}$ would be enough to
produce a small number of luminous carbon stars. In this grid of model
stars the $1.8\,{\rm M}_{\sun}$ star experiences a TDU event at the
last pulse. This TDU increases the carbon abundance slightly but does
not raise the ratio of C/O above 1. This is consistent with the
deficit of bright carbon stars in the bulge.  A single dredge-up would
be sufficient to bring an observable amount of Tc and other s-process
elements to the surface. Any carbon stars that form would do so when
the last pulse is close to the time when the envelope is ejected. The
smaller the mass of the envelope, the easier it is to pollute with
dredged-up material. The carbon star lifetime would be short in this
case and this would explain the lack of luminous carbon stars. The
inferred maximum mass would also not be sufficient to produce stars
which undergo HBB and/or SDU.

The inferred model value of [Fe/H] for the highest mass models is
+0.195. This is reasonably close to the maximum values found by
\citealt{ben10}. They found in their sample of bulge main sequence and
subgiant stars a bimodal distribution of [Fe/H], one with a low and
one with a high [Fe/H]. In their group with the highest values of
[Fe/H], the range of [Fe/H] is between +0.10 and +0.56 with a peak at
+0.3. The differences between the [Fe/H] found in this fit and
\citet{ben10} are small and can be reconciled by a different choice of
the level of the alpha plateau in this paper. In \citealt{ben11}
Figure~10 it appears the [O/Fe] plateau is closer to +0.5 then to
+0.35 as chosen by this paper. It is important to note that other
alpha elements are closer to the +0.35 plateau. With a higher plateau
value the downward slope of [O/Fe] is steeper. In this case for the
PNe models to match the O/H value a higher [Fe/H] value is required.

Another problem with this model is this produces an AGB star tip
luminosity ($M_{bol}=-5.14$) which is more luminous then observed
compared to what is seen in \citet{zo03} ($M_{bol}\approx-4.6$). This
is not surprising since TP-AGB stars are expected to follow a
core-mass luminosity relationship. As noted in
Subsection~\ref{sec:cspnmass} the maximum CSPN mass is probably less
than 0.62$\,{\rm M}_{\sun}$.

This model is in qualitative but not quantitative agreement with
observations of the bulge. It is in qualitative agreement because the
overall features of GB-PNe are reproduced, it matches the observation
of very few bright carbon stars, and has approximately the right
metallicity. It does not produce quantitative agreement since the
inferred AGB tip luminosity and progenitor masses are too high.

This model should be thought of as an upper limit to the bulge ZAMS
mass. The choice of He/H and O/H are as low as possible and the CSPN
mass is as high as possible. This choice of parameters produces the
largest possible ZAMS mass since the lower $Y$ is and the higher the
ZAMS mass, the larger the resulting CSPN mass.

\begin{figure*}
\vspace{54pt}
\includegraphics{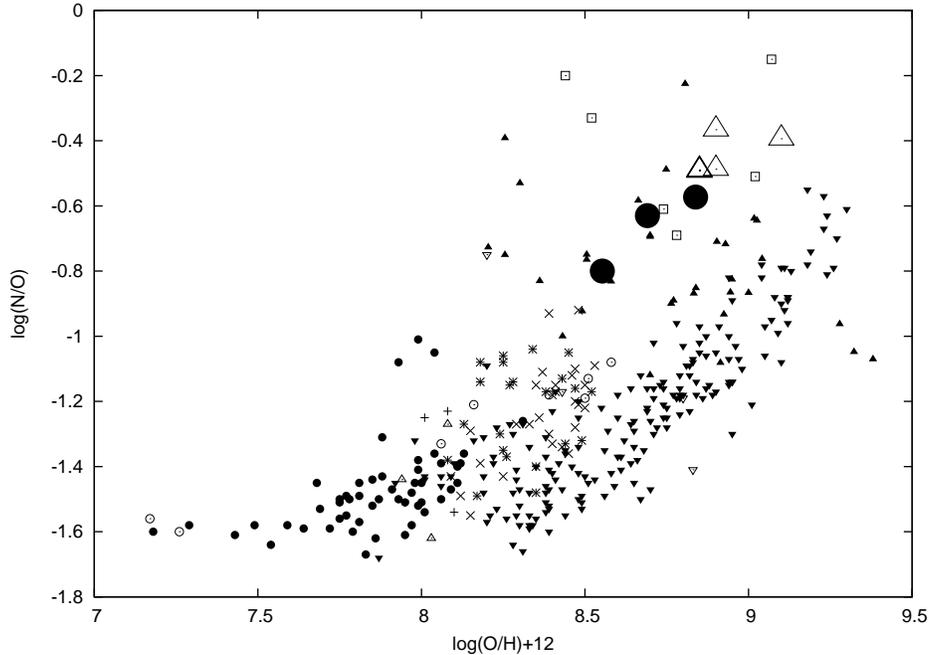}
\caption{Plotted in this figure is a diagram similar to Figure 1b in
  \citet{hek00} which shows the cosmic evolution of N/O as a function
  of O/H. The figure shows the inferred ZAMS O/H and N/O of both the
  Wang \& Liu fits and the less conservative fits. The large circles
  and triangles are the Wang \& Liu and less conservative fits,
  respectively. The pluses, Xs, stars, open circles, closed circles,
  open upright triangles, closed upright triangles are the HII region
  abundances of \citet{her09}, \citet{bre09}, \citet{mag07},
  \citet{gar99} and references therein, \citet{it99}, \citet{ks99},
  \citet{acw97}, \citet{fs91}, and \citet{vz98}, respectively. The
  open squares are the OB star abundances from \citet{tr02}.}
\label{fig:h2nooh}
\end{figure*}

An advantage of the Wang \& Liu fit is since it has well defined
trends it is possible to compare the inferred ZAMS N/O to the observed
chemical evolution of the Universe. In Table~\ref{tab:wangandliu} are
the typical parameters of less metal-rich and presumbably older ZAMS
stars are shown. Figure~\ref{fig:h2nooh} shows the cosmic evolution of
N/O as a function of O/H. The positions of both galactic and
extragalactic HII regions and some B stars are indicated. For
comparison purposes the inferred ZAMS O/H and N/O of the Wang \& Liu
fits (indicated by the large circles) are also plotted. The inferred
position of the GB-PNe on the N/O-O/H plane are comparable to the
positions of HII regions and unevolved stars. The chemical evolution
of nitrogen suggested by the GB-PNe suggests the chemical evolution of
the bulge is consistent with what is seen elsewhere in the
Universe. The GB-PNe show an upward slope of N/O as a function of O/H.
This is consistent with the upward slope of N/O versus O/H seen in HII
regions and stars. This trend is interpreted as indicating the
production of secondary nitrogen becoming important \citep{hek00}. The
conclusion is chemical evolution models of the bulge need to include
both primary and secondary production of nitrogen.

\subsection{Less conservative scenarios}
\label{sec:lc}

The largest ZAMS mass inferred from the Wang \& Liu fit is higher
than the MSTOs observed by studies of stars, therefore, models are fit
to a series of reasonable assumed parameters which will produce a
lower inferred MSTO mass. The sets of global parameters fit are listed
in Table~\ref{tab:lesscons}. These are fits only to the high mass end
of the CSPN mass distribution. The primary changes are a decrease in
the highest CSPN masses and in some cases an increase in the He/H
value. This should give lower ZAMS masses since a lower main sequence
mass should give a lower CSPN mass. Higher values of $Y$ on the main
sequence lead to larger CSPN masses as well also giving a lower ZAMS
mass.

Models are fit to GB-PNe He/H values of 0.130 to 0.140. In
Figure~\ref{fig:bulgeheolimited} the positions of these models are
indicated by the large triangles. The triangles seem to be near the
top of the He/H distribution as defined in
Subsection~\ref{sec:mohe}. The maximum value He/H is set to is 0.140
because this is the highest value which shows up in all of the GB-PNe
sets. The assumed value of O/H for most sets is between 8.85 or
8.90. Thes seems to agree well with the highest value found in all the
GB-PNe sets as indicated in Figure~\ref{fig:bulgeheolimited}. The
value of N/O were set to -0.200 to -0.300. Figure~\ref{fig:bulgeheno}
shows the position of the these fits which can be seen near the top of
all the distributions.

\begin{table*}
\caption{Parameters to fit for less conservative parameters}
\label{tab:lesscons}
\begin{tabular}{ccccccccccc}\hline 

CSPN mass&He/H&log(O/H)+12&log(N/O)&$M$&$Y_{ZAMS}$&$Z_{ZAMS}$&[Fe/H]&log(N/O)$_{ZAMS}$&$M_{\rm AGB-tip}$&Notes\\\hline

0.600&0.130&8.85&-0.300&1.60&0.320&0.0176&0.210&-0.491&-4.94&\\ 

0.600&0.140&8.85&-0.300&1.60&0.340&0.0186&0.195&-0.493&-4.97&\\ 

0.580&0.130&8.85&-0.300&1.40&0.319&0.0190&0.195&-0.490&-4.74&\\ 

0.590&0.130&8.85&-0.303&1.25&0.316&0.0190&0.205&-0.490&-4.80&(1)\\ 

0.590&0.130&9.10&-0.220&1.35&0.311&0.0363&0.575&-0.394&-4.91&(2)\\

0.580&0.135&8.90&-0.200&1.50&0.331&0.0221&0.270&-0.366&-4.74&(3),(4)\\

0.590&0.135&8.90&-0.300&1.35&0.325&0.0214&0.270&-0.487&-4.86&(1),(3),(4)\\
\end{tabular}
\smallskip

The CSPN mass, He/H, log(O/H)+12, and log(N/O) are the values of the
PN parameters being fit. $M$, $Y_{ZAMS}$, $Z_{ZAMS}$,[Fe/H] and
log(N/O)$_{ZAMS}$ are the model ZAMS parameters of the best fitting
model. $M_{\rm AGB-tip}$ is inferred bolometric magnitude of the
AGB-tip\\ (1) This model has its mixing length parameter increased to
2.2 from 1.8 which decreases TP-AGB mass-loss rate.\\ (2) This is a
model fit to a higher O/H value then is indicated by the PNe oxygen
abundances.\\ (3) In these models the pre-AGB mass loss one half of
that given by the formulas in Subsection~\ref{sec:rgbml}.\\ (4) In
these models the relationship between the abundances of the $\alpha$
elements and iron was modified. The value of [$\alpha$/Fe] plateau was
set to +0.5. This leads to a steeper decline in [O/Fe] as a function
of [Fe/H].
\end{table*}

The less conservative models listed in Table~\ref{tab:lesscons} are
reasonable fits to observations of GB-PNe. The parameters have been
chosen to cover a range of possible parameters of the GB-PNe with the
highest mass progenitors. They are considered less conservative since
the typically assumed CSPN mass is less than 0.62${\rm M}_{\sun}$
which will be fit by lower mass ZAMS models.  All the models have a
supersolar a ZAMS [Fe/H] between +0.2 and +0.3. This agrees with the
\citealt{ben10} observations of super solar bulge stars. As noted in
Subsection~\ref{sec:wl}, they find the [Fe/H] of the youngest stars
are between +0.15 and +0.5.

The first two models listed in Table~\ref{tab:lesscons} are chosen to
fit a maximum CSPN mass of 0.60${\rm M}_{\sun}$ with two different
values of He/H. The inferred progenitor masses of both are 1.6$\,{\rm
  M}_{\sun}$ progenitors. The inferred ages of both stars
$\approx2\,{\rm Gyr}$, which is younger than the youngest inferred
ages from \citet{ben10,ben11}. The model AGB tip magnitudes are. For
the first two models in Table~\ref{tab:lesscons} the bolometric
magnitudes are -4.94 and -4.97 which is close to the observed value
\citep{zo03}.

The third model listed in Table~\ref{tab:lesscons} was chosen to fit a
maximum CSPN mass of 0.58${\rm M}_{\sun}$. This is a CSPN mass as low
as is possible which fits the observations. This of course may be
too low. The ZAMS mass is 1.4$\,{\rm M}_{\sun}$ model has an age around
3.0\,{\rm Gyr} which agrees with the youngest star in the Bensby
work. The inferred $M_{AGB-tip}=-4.74$ is in good agreement with the
observed value.

None of the first three models listed in Table~\ref{tab:lesscons}
experience a TDU event although reasonable changes to the parameters
in the dredge-up law can give dredge-up events near the tip of the
AGB. This observation is consistent with the lack of bulge carbon
stars.

\subsection{Models with variations in the mass-loss laws}
\label{sec:ml}

One way to get the models into better numerical agreement the inferred
masses of \citet{ben10,ben11} would be to decrease the mass-loss on
the TP-AGB. This gives longer TP-AGB lifetimes, allowing the core more
time to grow to a higher mass. There are a number of ways of doing
this to reduce the mass loss rates, but the way chosen for this study
is to increase the mixing length parameter, $\alpha$. Increasing
$\alpha$ makes the star more compact and leads to a decrease in the
mass-loss rate. The models where this procedure has been applied are
indicated in rows four and seven of Table~\ref{tab:lesscons}.

In the model in row four the best fitting model is a 1.25$\,{\rm
  M}_{\sun}$ model with slightly higher than solar abundances. This
model gives a CSPN mass of 0.59$\,{\rm M}_{\sun}$ and indicates the
bolometric magnitude of the AGB tip is -4.8 which is in pretty good
agreement with the observed value. The pre-TP-AGB lifetime of this
model is around 4-5 Gyr. The ZAMS helium abundance of this model is an
elevated to $\approx0.32$. 

In the model in row seven the best fitting model is a 1.35$\,{\rm
  M}_{\sun}$ model with super-solar abundances. In this model the
relationship between the ZAMS abundances of the $\alpha$ elements and
the abundance of iron has been modified. The plateau has been
increased to +0.5. This was chosen since when \citet{ben10} plotted
[O/Fe] as a function of [Fe/H] the plateau for oxygen (and only
oxygen) is at +0.5. What this means is [O/Fe] has a steeper decrease
as a function of [Fe/H] for [Fe/H]$>$-1. This means to get the same
O/H a higher metallicity is required. This model is probably the best
numerical match to the Bensby et al. observations since the implied
[Fe/H] of +0.27 and the inferred highest mass progenitor matches their
highest mass closely.

In the sixth model listed in Table~\ref{tab:lesscons} the pre-AGB mass
loss was reduced to one half the value as determined by the procedure
outlined in Subsection~\ref{sec:rgbml}. The result is not
significantly different then if the pre-AGB mass loss were calculated
normally.

\subsection{A model which fits a higher oxygen abundance}
\label{sec:ho}

As noted in Section~3 there is a possibility the oxygen abundance
might be higher than is indicated by the PN abundances. The PN
parameters fit and the ZAMS parameters used to fit it are indicated in
Table~\ref{tab:lesscons}. With an nearly double oxygen abundance
this has a much higher value of $Z$, which is closer to the Bensby
results than the other results. The mass of the fit is $1.35\,{\rm
  M}_{\sun}$, which is also close to the Bensby result. The ZAMS value
of $Y$ is $\approx0.31$.

\subsection{Summary of models}

All of the models above are reasonable fits to the GB-PNe trends. They
all produce CSPN masses and AGB tip luminosities comparable to the
observations. None of them would produce luminous carbon stars except
via blue stragglers (not modeled in this paper). All of the implied
ZAMS abundances are comparable to what is seen elsewhere. All of the
models consistently require an ZAMS value of $Y$ between 0.31 and
0.34. The conclusion is the highest mass bulge ZAMS stars should have
$Y$ in this range. The inferred ZAMS N/O ratios are greater than the
solar N/O. In all models the ZAMS C/O is less than 1. This agrees with
the lack of bright carbon stars in the bulge. More observations of
bulge C/O are needed to quantitatively constrain carbon.

\section{Discussion}
\label{sec:diss}
From these models it is concluded it is possible, with reasonable
choices of input parameters, to get a qualitative agreement with the
observations. All the different assumed scenarios result in
essentially no luminous carbon stars which matches the observations. The
ZAMS masses and ages needed to reproduce the CSPN masses are
reasonably close to the inferred ZAMS masses of
\citet{ben10,ben11}. The model AGB tip luminosities are in good
agreement with the measured luminosities of \citet{zo03}.

If these models are correct they indicate that the element helium had
a different chemical evolution in the bulge compared to its evolution
in the disc. With the exception of the high oxygen abundance model,
for the models to reproduce these GB-PNe the ZAMS values of $Y$ are
consistently $\sim0.32$. The $Z$ values needed to fit the most massive
models range from $\sim0.019$ to $0.22$. If a primordial values of the
helium abundance, $Y_0$, between 0.24-0.25 are assumed this means for
the bulge $\frac{dY}{dZ}\approx4$. This is significantly higher than
the typical values of 1.4-1.8 which are typically quoted. Using a
typical values of $\frac{dY}{dZ}$ and $Y_0$ a star with a $Z=0.019$
should have values of $Y$ between 0.26 and 0.29 indicating. The model
with a higher oxygen abundance also required an elevated $Y$ of 0.31
to produce the highest He/H values of GB-PNe.

In all of the models of the highest mass progenitors, both the Wang \&
Liu and the less conservative models, the ZAMS ratio of N/O had to be
increased above the solar ratio to match the observed GB-PNe N/O. To
match the typical nitrogen abundances it was necessary to enhance the
nitrogen abundance. The typical ZAMS N/O ratio was between 1/4 and 1/3
which is higher than the ratio for the Sun ($\sim1/9$). This is
approximately a factor of 3 enhancement over the solar ratio.

All of this suggests that the younger bulge stars have been enhanced
by stars ejecting the products of the CNO cycle. As shown in
figure~\ref{fig:h2nooh} the ZAMS N/O for the less conservative and the
Wang \& Liu scenarios is consistent with bulge chemical evolution
scenarios where nitrogen is produced as both a primary and a secondary
element. As a secondary element the nitrogen yield is dependent on
metallicity. It is not surprising then that the starting N/O of these
stars is higher than solar.

Where is the extra helium synthesized? One possibility for the site of
this production is intermediate-mass stars. \citealt{ben10,ben11}
found a bimodal distribution in [Fe/H] and the age of the microlensed
stars. One of the peaks contains stars with subsolar [Fe/H] and
typical ages of 10-12 Gyrs. The other peak contained stars with [Fe/H]
typically between 0 and +0.5 and younger ages ranging from 3-7 Gyrs in
age. There appears to be a gap in the distribution of [Fe/H]
indicating star formation stopped for a time. The older population
would have formed quickly. Because of this quick formation would have
incorporated the products of supernovae of type II and not that of
type I supernova and intermediate-mass stars. The high-mass and
intermediate-mass members of this older population would then have
polluted the interstellar medium (ISM) with helium-rich material. If
the existing ISM were very low mass this could potentially lead to an
over-enrichment of helium.

\subsection{Can the observations be reproduced with intermediate-mass stars?}

\citet{mar03} compared synthetic TP-AGB models to the abundances of
PNe. For \citealt{mar03} to produce the high values of He/H near 0.13
stars of mass $\ga3.5\,{\rm M}_{\sun}$ were required. A star of this
mass would, according to the AGB tracks in \citet{bertb}, produce a
CSPN of 0.72$\,{\rm M}_{\sun}$ and an AGB tip magnitude of -5.7. This
does not match the observations, and unless the galactic centre is
significantly farther away then believed, this scenario is
excluded. Also if this intermediate-mass star scenario is correct,
there would be stars in the bulge of the right mass to produce bright
carbon stars, this is not observed in the bulge. All of this together
suggests that stars massive ($>2.0\,{\rm M}_{\sun}$) enough to produce
the enhanced helium abundance do not exist in large numbers in the
bulge. Therefore, the intermediate-mass star scenario is ruled out.

\subsection{Can these observations be reproduced by a very low-mass star?}

To test the question if a star with an age of $\sim10\,{\rm Gyr}$
could reproduce the observations a lower mass model has been run. This
is the approximate age found by \citet{zo03} for the galactic
bulge. For $Y\approx0.30$ this would require a star with main sequence
mass $\sim1\,{\rm M}_{\sun}$. The model parameters were adjusted to
fit He/H=0.130 and $\log{(O/H)}+12=8.85$. A reasonable model is found
with M$_{\rm ZAMS}$=1.05$\,{\rm M}_{\sun}$, [Fe/H]=0.19, $Y=0.314$ and
$Z=0.0199$. However, the CSPN mass of this is only 0.56$\,{\rm
  M}_{\sun}$. This is lower than the observed maximum mass CSPN in the
Galactic bulge.

\section{Conclusions and Recomendations for future work}
\label{sec:con}

This study has shown
\begin{enumerate}
\item{Reasonable TP-AGB models with low-mass progenitors
  (1.2-1.8$\,{\rm M}_{\sun}$) and with enhanced $Y$ and N/O on the
  ZAMS can at least qualitatively match the observations of the
  GB-PNe. This match includes the observed values in GB-PNe of He/H,
  O/H, N/O and the CSPN mass. These models also qualitatively match
  the bulge's AGB tip luminosity and are consistent with little or no
  formation of bright carbon stars.}
\item{The inferred N/O values for the progenitor ZAMS stars are
  consistent with what is known about the cosmic evolution of
  nitrogen. The implication is the secondary production of nitrogen is
  important in the chemical evolution of the Galactic bulge.}
\item{Very low-mass (M$\approx1\,{\rm M}_{\sun}$) and
  intermediate-mass (M$\ga3.5\,{\rm M}_{\sun}$) are not able to match
  the observed CSPN masses. Unless the CSPN masses are very different
  from what is observed the bulge, the very low- and intermediate-mass
  stars would produce CSPN masses that are either too low or too high,
  respectively. The intermediate-mass star scenario would produce very
  bright carbon stars which are not observed.}
\item{The youngest stars in the bulge are probably between 2 and 5
  Gyrs in age and have a slightly higher metallicity than the Sun
  ($Z\approx0.019$ to $0.022$) and a significantly higher value of $Y$
  ($\approx0.32$ to $0.34$) than a disc star of the same
  metallicity. This is in rough agreement with the results of
  \citet{ben10,ben11} for bulge main sequence and subgiant branch
  stars.}
\item{These models indicate the last galactic bulge stars formed
  between 2 and 4 Gyrs of age. This should be thought of as an
  intermediate age population.}
\item{The implied C/O of most of the GB-PNe is less than 1. This
  requires a ZAMS C/O$<1$.}
\end{enumerate}

Some recommendations for future work include: Better values of the
CSPN mass need to be determined. A big limitation of this study is the
poor knowledge of the CSPN masses so any improvements would lead to
better limits on the range of possible models. Future microlensing
studies of bulge stars should include carbon and nitrogen abundances
if possible. This paper predicts the ratio of N/O of super-solar stars
should be enhanced relative to the solar values.

\section*{Acknowledgments}
I would like to acknowledge the support of a Alfred State College
Faculty Scholarship Grant which supported this work during the Spring
2012 semester. I would also like to thank R.B.C. Henry for reading an
earlier manuscript and making comments which have improved this
paper. I would like to acknowledge the contribution of the papers
referee, D.M. Nataf, whose comments considerably improved the paper.

\label{lastpage}

\end{document}